\documentclass{article}[11pt]
\usepackage[utf8]{inputenc}
\usepackage{cite}
\usepackage{graphicx}
\usepackage{subcaption}
\usepackage[margin=1in]{geometry}
\usepackage{amsmath}
\usepackage{array}
\newcolumntype{P}[1]{>{\centering\arraybackslash}p{#1}}
\newcolumntype{M}[1]{>{\centering\arraybackslash}m{#1}}
\title{
Meta-Analytic Operation of Threshold-independent Filtering (MOTiF)
Reveals Sub-threshold Genomic Robustness in Trisomy
}
\author{ Roy Siegelmann \\  rsiege15@jhu.edu \\
        Department of Applied Mathematics and Statistics\\
       Johns Hopkins University, \\
       Baltimore, MD 21218-2680, USA \\ \ 
       \and
         Hava T. Siegelmann \\  hava@cs.umass.edu \\
        School of Computer and Information Sciences\\
       University of Massachusetts, Amherst\\
       Amherst, MA 01003-9264, USA
    }

\date{}

\begin{document}

\maketitle

\begin{abstract}

Trisomy, a form of aneuploidy wherein the cell possesses an additional copy of a specific chromosome, exhibits a high correlation with cancer. Studies from across different hosts, cell-lines, and labs into the cellular effects induced by aneuploidy have conflicted, ranging from small, chaotic global changes to large instances of either overexpression or underexpression throughout the trisomic chromosome. We ascertained that conflicting findings may be correct but miss the overarching ground truth due to careless use of thresholds. To correct this deficiency, we introduce the Meta-analytic Operation of Threshold-independent Filtering (MOTiF) method, which begins by providing a panoramic view of all thresholds, transforms the data to eliminate the effects accounted for by known mechanisms, and then reconstructs an explanation of the mechanisms that underly the difference between the baseline and the uncharacterized effects observed. As a proof of concept, we applied MOTiF to human colonic epithelial cells, discovering a uniform decrease in gene expression levels throughout the genome, which while significant, is beneath most common thresholds. Using Hi-C data we identified the structural correlate, wherein the physical genomic architecture condenses, compactifying in a uniform, genome-wide manner, which we hypothesize is a robustness mechanism counteracting the addition of a chromosome.  We were able to decompose the gene expression alterations into three overlapping mechanisms: the raw chromosome content, the genomic compartmentalization, and the global structural condensation. While further studies must be conducted to corroborate the hypothesized robustness mechanism, MOTiF presents a useful meta-analytic tool in the realm of gene expression and beyond.
    
\end{abstract}

\section{Introduction}

Trisomy is a common cellular mutation, occurring in more than 1\% of pregnancies and a similar percentage of somatic cellular divisions. However, trisomic cells almost universally undergo apoptosis, and gametes with the mutation rarely achieve viability, either spontaneously aborting or resulting in stillbirths. Within trisomic cells that did not undergo apoptosis, there is a high correlation with cancerous growth, as more than 90\% of solid tumors possess some sort of aneuploidy, along with nearly 85\% of haematopoietic cancers. \cite{1,2,3,4,5,6,7,8,9} Aneuploidy can be observed even in premalignant carcinomatous cells, and its presence can be seen as an indicator for future tumor development, wherein trisomy-7 is among the earliest observed. \cite{10,11,12,13,14,15,16,17,18}. It is well documented that the addition of an extra chromosome not only precipitates an increase in gene expression of the trisomic chromosome, but also a global disjunctive loss of homeostatic regulatory patterns throughout the genome, referred to as ‘transcriptional dysregulation’\cite{19}, ‘genomic instability’\cite{20}, ‘genome chaos’\cite{21}, and an ‘imbalance in the levels of gene product’\cite{22}.  

One line of research proposes that the addition of a chromosome (a 50\% increase in local genetic material) causes increased gene expression levels approaching 50\%, which in turn leads to aneuploidy’s observed phenotypic variation. \cite{5,23,24} This differs from other findings suggesting a global effect, wherein 80–95\% of misregulated genes are not contained on the trisomic chromosome. \cite{19} Similarly, another study reported that out of 79 proteins which differentially expressed in manufactured trisomic cells, not one was found on the trisomic chromosomes. \cite{25} Interestingly, a study focusing on aneuplodic plants reveals only a small increase in gene expression on the trisomic chromosome, and a broad decrease throughout the remainder of the genome. \cite{26,27} This stands in stark contrast to the majority of previous findings, which indicated a positive correlation between gene expression levels and copy number. \cite{28} 

We start by demonstrating that the majority of differences in previous studies can be explained by methodology of analysis, of which perhaps the most common method is to examine the percentage of genes (or segments of the chromosome) which increase (or decrease) by at least a certain threshold. The selection of different thresholds has already been a point of contention in another research area, wherein a clash about whether the number of genes in metastatic cells on the chromosomes with an additional arm increasing expression was 3.8\% or above 80\% was resolved by altering the selected threshold. \cite{24,29} Percentage thresholding is potentially useful for focusing on specific magnitudes of change, but lacks the sort of global view which is necessary to fully understand the effects of aneuploidy. Instead of using specific thresholding, we suggest using the full panorama of thresholds followed by an analysis which is entirely threshold independent. That is, to look at the full panorama of possible thresholds and considering each recognized effect of biological relevance, using them to gain a holistic threshold independent understanding.  This is the at the heart of the new meta-analytic method we propose, which we call the Meta-analytic Operation of Threshold-independent Filtering (MOTiF) method, wherein the gene expression data is renormalized by successively removing previously identified effects until either reaching the expected baseline or finding effects that were heretofore undiscovered.

As a model cell line for analysis, we used human colonic epithelial cells (HCEC) and compared these to an analogous cell line to which trisomy-7 was induced (HCEC+7). \cite{30,31} Future studies will utilize MOTiF for analysis on other hosts and aneuploidies. Applying MOTiF to these cell lines, we noted that while the original, raw results may imply a positive correlation between gene expression and copy number, another effect compounded upon this base. This effect modifies gene expression levels globally, such that all chromosomes display a relatively uniform decrease in expression levels, acting upon the 50\% increase in expression on the trisomic chromosome (from copy number increase) to provide a complex, multifaceted final gene expression product. We analyzed the structural component by applying graph-theoretic analysis to the Hi-C data and found a significant condensation in the physical network structure of the genome. These manufactured aneuploidic cells exhibit a small yet significant decrease in the radius of the network, a striking increase in the density of the network, and larger densely-connected areas, yet their overarching global structure is unchanged. These architectural shifts are uniform throughout all chromosomes and throughout the entirety of each chromosome. Together with our observed gene expression alterations, they support the hypothesis that these changes are the byproducts of a genome-wide cellular robustness mechanism used to counteract the most pernicious effects caused by the addition of a chromosome - namely the change in nuclear density and gene expression levels.

This effect is the missing link necessary to properly disassemble the complex dysregulation pattern of aneuploidic gene expression into three component parts. These are the additional chromosome (yielding a raw 50\% increase in Chromosome 7 gene expression), the complex A-B compartment shifting that is responsible for the transcriptional dysregulation observed at high thresholds, and the global chromosomal condensing effect proposed here, which has previously gone undetected due to its pervasive sub-threshold effects. An understanding of the entire picture will enable us to understand the nature of cellular chromosomal aberrations more completely. Due to the strong correlation with pre-carcinomatous cell lines, the identification of new effects as an indication to the presence of aneuploidies may provide crucial insight into the development and maintenance of carcinomatous cell lines and tumors. While the mechanism of the global genomic condensation remains a hypothesis due to limited data, we introduced the MOTiF method and demonstrated its capabilities. Learning to properly use this multifaceted meta-analysis will advance gene expression research to  new heights, wherein past discoveries  become the building blocks upon which new discoveries are made.

\section{Results: Differential Gene Expression}

We begin by providing a comprehensive view of the functional genomic landscape. Our gene expression data is taken from normalized and cleaned RNASeq vectors, wherein each chromosome which represents the parts-per-million of mRNA fragments that come from specific segments of DNA. To receive the gene expression per length as opposed to pure gene expression, we normalized the data by dividing the aneuploidic chromosome by its aneuploid factor. In this case, we divide the values for Chromosome 7 by 3/2 since there are now three copies, as opposed to the two normally present. This has the effect of counteracting the effect caused by the presence of a third copy of Chromosome 7, which is our first application of MOTiF, leaving the effects on gene expression which cannot be explained solely by the quantity of genetic material. The average gene expression level throughout the genome decreases significantly from the HCEC to the HCEC+7 cell line (Figure 1a), with nearly all chromosomes lowering gene expression levels and displaying expressional dysregulation (see Table 2 for statistical significance). Chromosome 7 increases by approximately 18\% in the raw data, however when normalized to mitigate the effects of increased genetic material, exhibits a significant decrease in gene-expression-per-length, akin to all others (Figure 1b). The distributions of gene expression shift from HCEC to HCEC+7 are similar throughout all non-trisomic chromosomes and the length-corrected trisomic chromosome, with a uniform decrease in the median value of gene expression change, and a left-skewed distribution (Figure 1c). 

To analyze the changes occurring throughout the genome, we must look beyond the global gene expression and examine how different segments of the chromosomal landscape vary their respective expression levels. Following state of the art analysis, we first examine the percentage length of each chromosome differentially expressing positively and negatively based on fixed thresholds (Figure 2a,b) before adopting a view free of specific thresholding and creating the full panorama which we will utilize for MOTiF (2c,d). To demonstrate the dangers of thresholding, we consider the effects of using 1.2 versus 2. In this case, an average of 28\% of each chromosome decreased by a factor of at least 1.2, whereas only 15\% increased by a similar factor (Figure 2a). With a threshold of 2, only an average of 8.94\% of each chromosome increased by a factor of 2 and 9.80\% decreased by such a factor (Figure 2b). Strikingly, in both cases, all chromosomes had similar percentages for increase and decreases, and Chromosome 7 is well within the range of other chromosomes.

As selecting a specific threshold alters the analysis of expressional changes, we provide a full panoramic landscape of the average increase and decrease percentages based on differential expression thresholds ranging from 1 to 11 by steps of 0.1 (Figure 2c). For all thresholds, there is a low standard deviation across chromosomes, and thus we use the average as our metric of choice.  One might naively think that a high threshold such as 2 would be preferred, as it is less prone to mechanical noise. However, such a selection would lead to the loss of all biological processes which do not adhere to the arbitrarily chosen threshold, and in fact any individual threshold in isolation is insufficient for completeness. Figure 2c hints at two different components of the differential gene expression: one which is relatively constant and impacts small, select areas very strongly; the other decays inversely with the threshold and impacts most of the genome with a smaller change in expression. Whereas the former - which we will call Component 1 - is nearly balanced between increase and decrease, the latter - Component 2 - overwhelmingly affects a global decrease in gene expression, corresponding to the observed 10.66\% decrease in global gene expression levels of Figure 1b. Component 2 is most prevalent in the 1-1.5 threshold range, and the larger thresholds depict mostly Component 1 (for example, Figure 2a depicts both components, while Figure 2b depicts mostly Component 1). Figure 2d reinterprets the data as a histogram, wherein each threshold (1.1-1.2, 1.2-1.3, etc.) is one bar, and thresholds above 11 are combined. This visualization clearly accentuates the demarcation between Component 2 (on the left) and Component 1 (on the right). Component 2 is responsible for the smaller expressional shift which occur in broad swaths of the genome, while Component 1 yields massive shifts in select parts of the chromosomes.

\begin{figure}[htbp]
        \centering
        \begin{subfigure}[b]{0.475\textwidth}  
            \centering 
            \includegraphics[width=\textwidth]{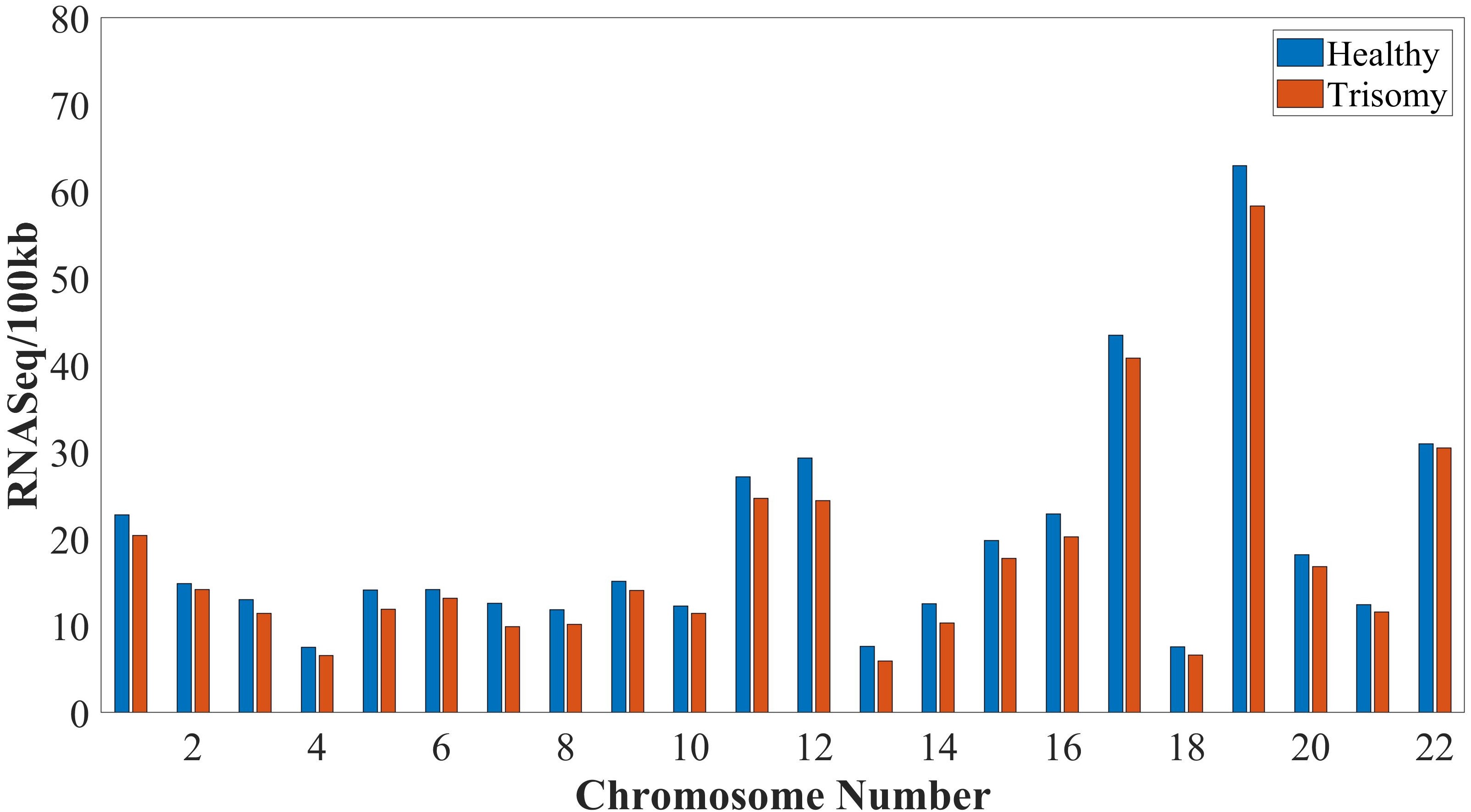}
            \caption[]%
            {{\small RNASeq gene expression per length in each chromosome}}    
            \label{fig:1a}
        \end{subfigure}
        \quad
        \begin{subfigure}[b]{0.475\textwidth}   
            \centering 
            \includegraphics[width=\textwidth]{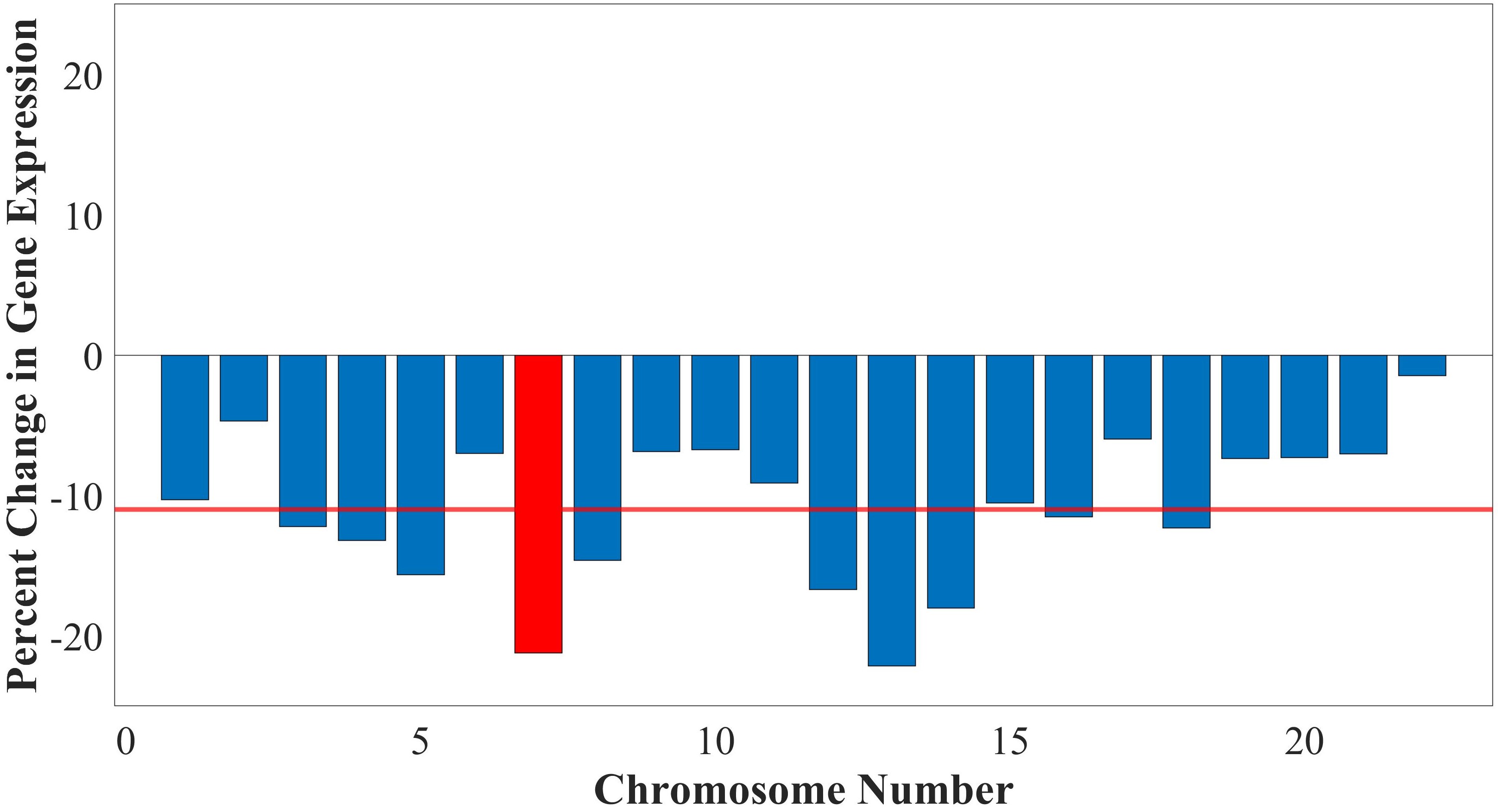}
            \caption[]%
            {{\small  Percent change in RNASeq expression per length in each chromosome (average of $10.66\%$ decrease)}}  
            \label{fig:1b}
        \end{subfigure}
        \vskip\baselineskip
        \begin{subfigure}[b]{0.95\textwidth}   
            \centering 
            \includegraphics[width=\textwidth]{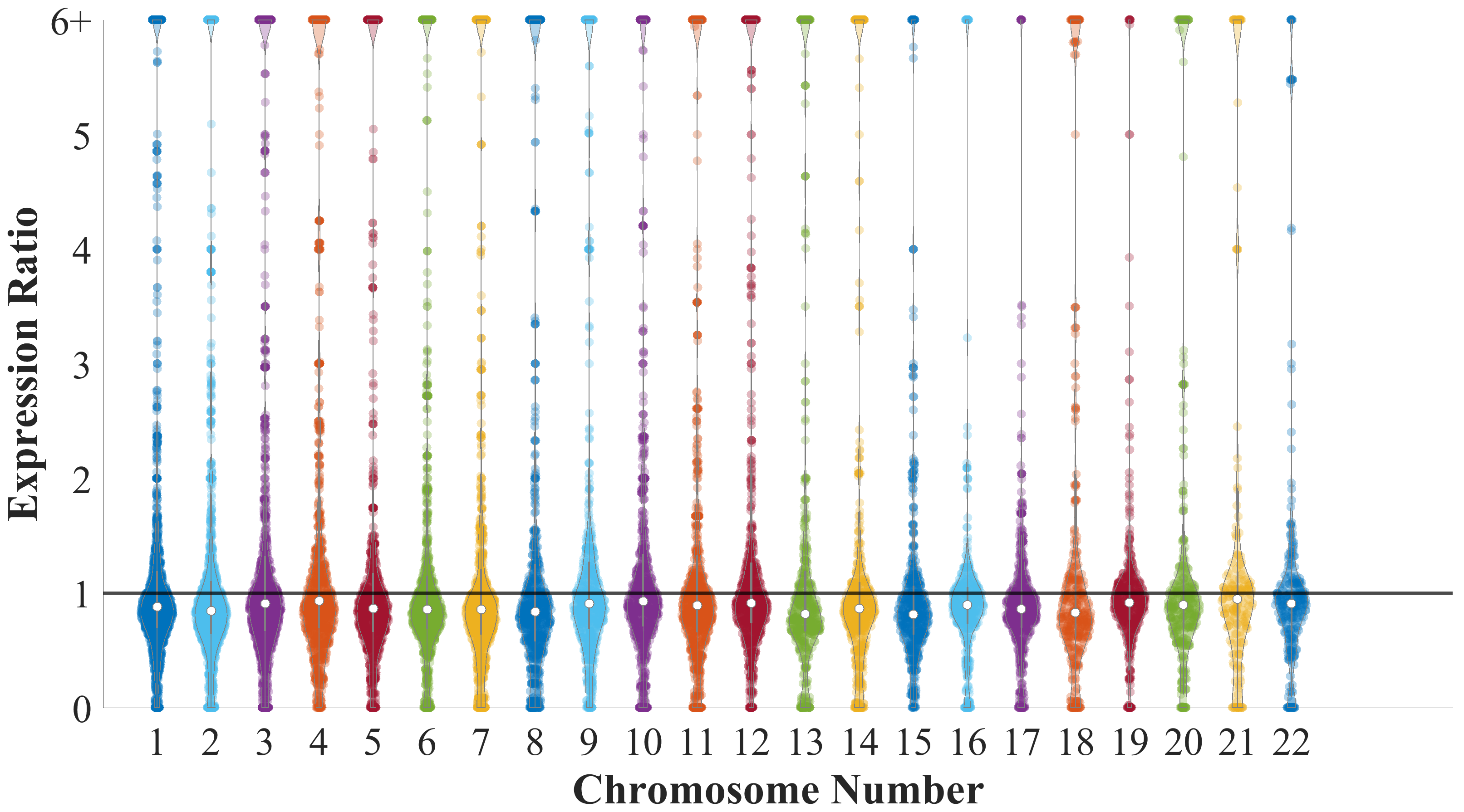}
            \caption[]%
            {{\small Distribution in expression change ratio per chromosome}}  
            \label{fig:1c}
        \end{subfigure}
        \caption{\textbf{Changes in gene expression throughout the genome from healthy to trisomy-7 variant.}  (a) A normalized average expression decreases in all chromosomes from the healthy (blue) to aberrant (red) cell lines. (b) The normalized percent decrease of gene expression in each chromosome shows that Chromosome 7 activity has among the greatest decreases (21$\%$, red bar), while other chromosomes decrease by about 10.66$\%$ on average (red line), all decreasing. (c) A violin plot containing the ratio change distribution of RNASeq in Trisomy divided by RNASeq in Healthy in each chromosome. The white dot in each chromosome depicts the median, which is lower than the baseline (black line, representing a ratio of 1) in all chromosomes. Everything above 6-fold increase is accumulated to the final point. As the distribution is centered below the baseline, every chromosome has more segments that decrease than increase in gene expression value. } 
        \label{fig:1}
    \end{figure}

  \begin{figure}[htpb]
        \centering
        \begin{subfigure}[b]{0.475\textwidth}
            \centering
            \includegraphics[width=\textwidth]{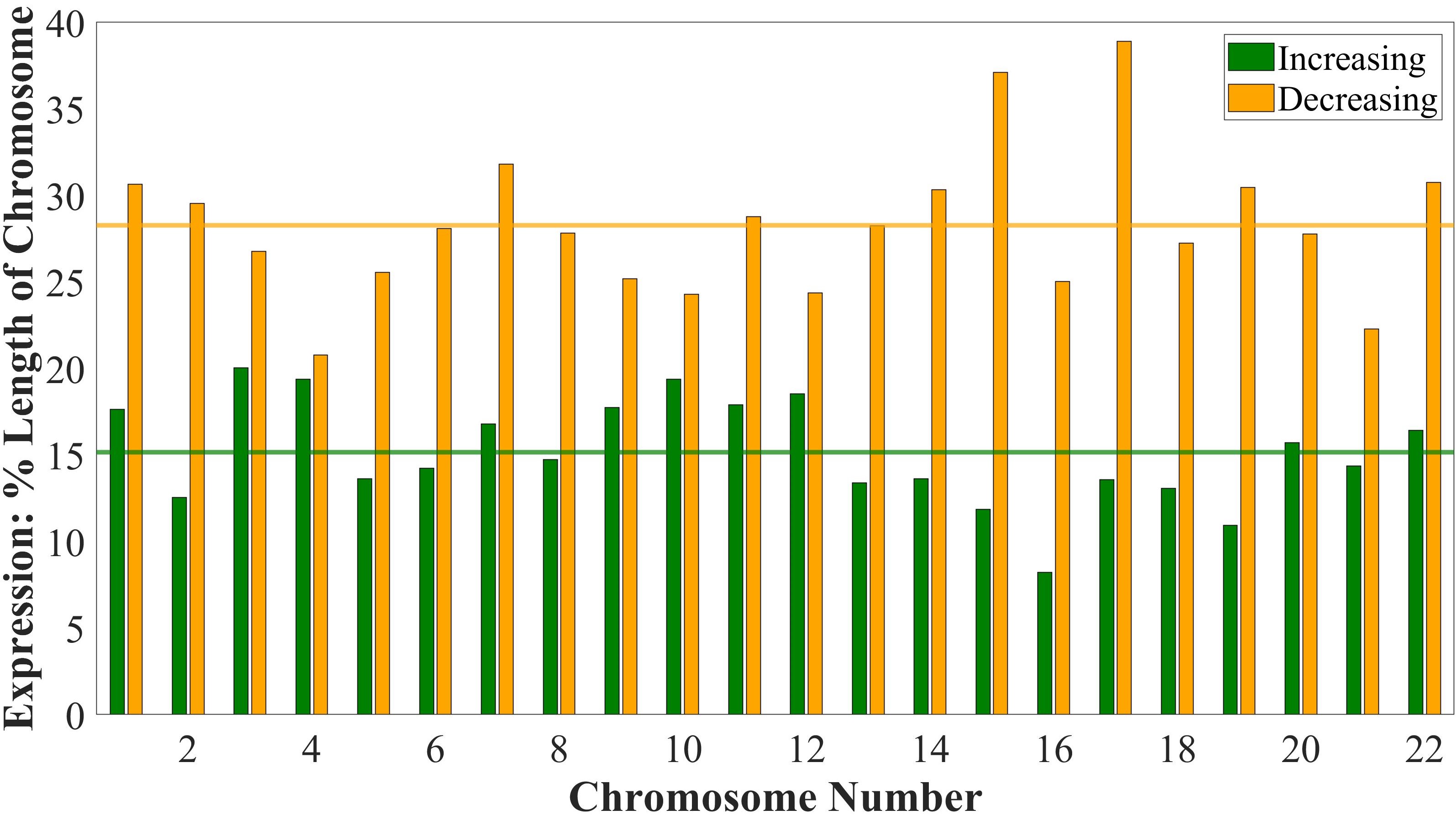}
            \caption[Network2]%
            {{\small Percent length per chromosome varying gene expression above a $1.2$x threshold. An average of $15.17\%$ increasing, $28.26\%$ decreasing}}    
            \label{fig:2a}
        \end{subfigure}
        \quad
        \begin{subfigure}[b]{0.475\textwidth}  
            \centering 
            \includegraphics[width=\textwidth]{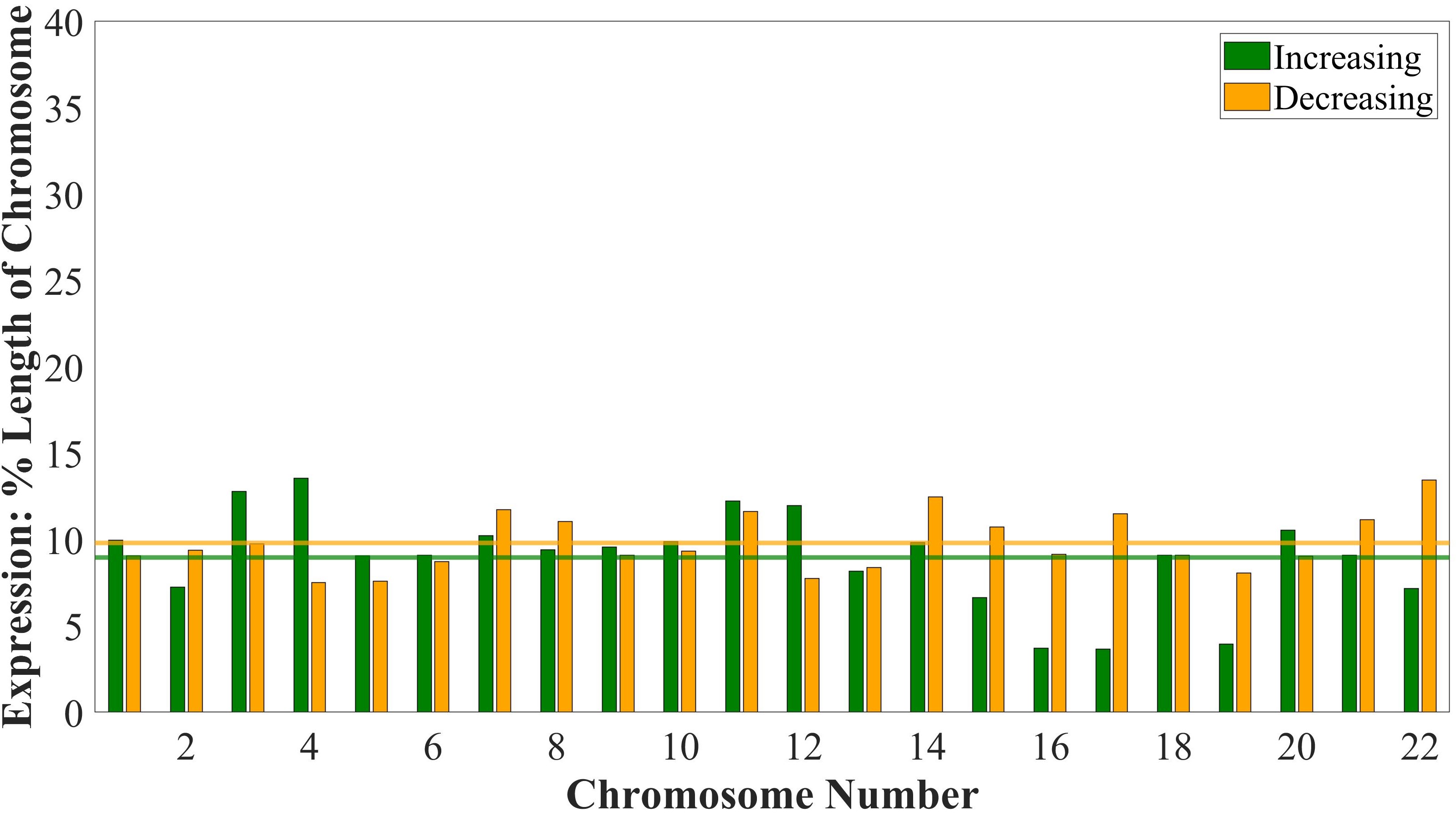}
            \caption[]%
            {{\small Percent length per chromosome varying gene expression above a $2.0$x threshold. An average of $8.94\%$ increasing, $9.80\%$ decreasing}}    
            \label{fig:2b}
        \end{subfigure}
        \vskip\baselineskip
        \begin{subfigure}[b]{0.475\textwidth}   
            \centering 
            \includegraphics[width=\textwidth]{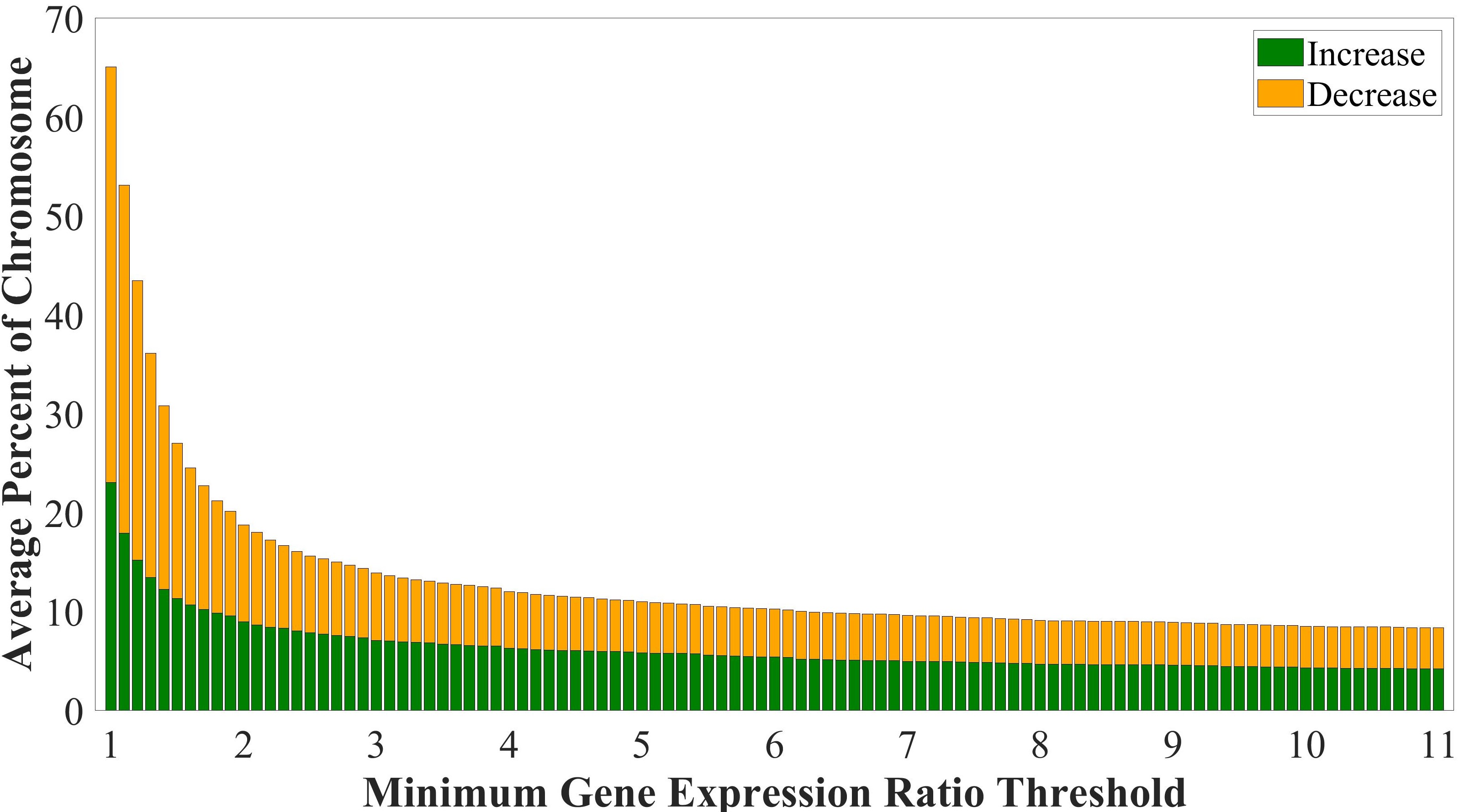}
            \caption[]%
            {{\small Average percent length per chromosome varying gene expression above threshold}} 
            \label{fig:2c}
        \end{subfigure}
        \quad
        \begin{subfigure}[b]{0.475\textwidth}   
            \centering 
            \includegraphics[width=\textwidth]{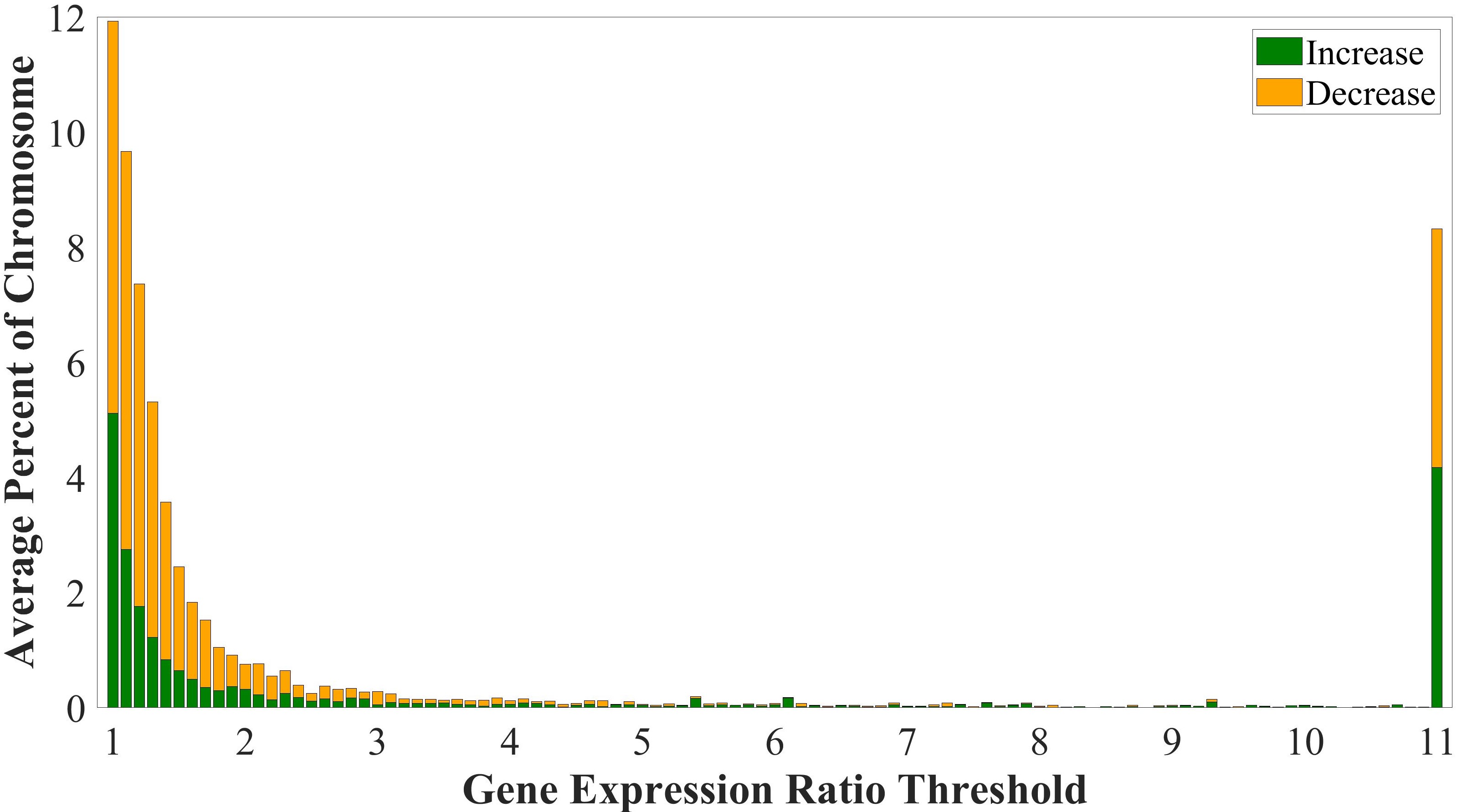}
            \caption[]%
            {{\small Average percent length per chromosome varying gene expression by exact threshold}}  
            \label{fig:2d}
        \end{subfigure}
        \caption{\textbf{Percent length of changes in gene expression from healthy to trisomy-7 variant.} (a) The percentage length of each chromosome whose gene expression changes above a threshold of $1.2$, i.e. either the trisomic expression divided by the wild-type expression for that segment is greater than $1.2$ (increase), or the wild-type divided by the trisomic is greater than $1.2$ (decrease). The green bar represents the percent increasing for each chromosome, the orange bar represents the percent decreasing for each chromosome, and the lines represent the genomewide means based on the color. Significantly more of the genome increases than decreases by these amounts. (b) Analog of part (a), with a factor of 2. Here, similar amounts of the genome increase and decrease above the threshold. (c) An expansion of (a) and (b) displaying the mean lines for each different threshold from 1 to 11 reveals a small constant percentage of the genome increasing by a factor of more than $6$, with slightly more increasing than decreasing, while in the smaller thresholds there is a strong preference for decrease. (d) The marginal variant of part (c), wherein only changes in exactly that location (i.e. between $1.1-1.2, 1.2-1.3$, etc.) are shown, such that the integral from this point to the end would correspond to the analogous bar in part (c), and the final bar shows everything that increases above the 11 threshold. The changes are much more significant toward the beginning, with a strong preference for decrease, while larger thresholds values exhibit small percentages in either direction.} 
        \label{fig:2}
    \end{figure}
    
 \newpage
 
\section{Results: Genomic Architecture}

Changes in gene expression typically result from changes in the genomic architecture - primarily through the folding and coiling of the DNA strand - as expression is lowered from increasing difficulty of access to the strand. This is demonstrated by chromatin packing levels, i.e., euchromatin being highly expressed and lightly condensed, heterochromatin being highly condensed and barely expressed. Euchromatin and heterochromatin play an important role in the high-level compartmentalization of DNA, wherein the A-compartment corresponds to gene-rich, highly expressed chromatin (with a high percentage of euchromatin), and the B-compartment corresponds to the densely packed less-expressed chromatin (with a high percentage of heterochromatin). \cite{32,33} Examination of A/B compartmentalization, and of the physical network architecture more generally, relies of the usage of Hi-C data.

Hi-C is a method of ascertaining the structural connectivity of the genome via a strength-adjacency matrix for the physical connections between different DNA stretches, with dimensions depending on the resolution. \cite{34} This matrix is symmetric, wherein each entry represents the strength of connectivity (and to some degree, the physical proximity) between the row segment and the column segment. The global Hi-C network before and after the induction of trisomy is visualized in Figure 4a. In both cell lines, the main diagonal - representing the connections between any given length of DNA and itself - is stronger than other entries. The small highly-connected blocks around the diagonal correspond to chromosomes. The trisomic variant possesses a bright cross-pattern around the seventh block, indicating the existence of an additional copy of that chromosome. Zooming in, we examine the Hi-C matrices of specific chromosomes in Figure 4b, which are similar to the global Hi-C network but have a higher resolution. 

Compartment shifting, with genes either ‘active’ in the A-compartment or ‘silenced’ in the B-compartment, appears to align well with Component 1. To validate that this is indeed the mechanism which explains Component 1, we examine A/B compartmentalization shifts from the wild-type to trisomic variants of the cells. Compartment shifts are around 4.5-5.5\% of the genome in each direction, with a slight preference for B-to-A (see Figure 3a). This aligns with our gene expression findings, as length and directionality of compartmentalization shifts (Figure 3a) match the length and directionality of chromosomal segments where gene expression changes by eleven-fold or more (Figure 2c). According to MOTiF, we have explained away Component 1 as the functional analogue to A/B compartmentalization, and now need to continue searching for the structural analogue of Component 2.

We next measure the algebraic connectivity (for details see Appendix B) in each chromosome, as an approximate of its physical density. The percent change in algebraic connectivity from the HCEC to HCEC+7 cell lines is drastic, averaging approximately 39\% (Figure 3b). This implies that there is a significant increase in intra-chromosomal connectivity throughout the genome. Another measure of connectivity is provided by TADs. TADs (Topologically Associating Domains) groups of chromosomal segments which all associate strongly with each other (see Figure 4b for Chromosome 7, Figure S1 for all others).  Genes in these regions generally share enhancers and promoters, change expression in a similar fashion, and are part of the same compartment as other genes within the same domain. \cite{35}

From the HCEC cell line to the HCEC+7 cell line, we observe changes in TAD size and number, with an average of 11.35\% increase in the mean size of TADs and 9.59\% decrease in the number of TADs, with nearly all chromosomes shifting in the same direction (Figure 4c). TAD size and number changing inversely signifies that the entire chromosomal network is becoming increasingly compactified, indicating a structural component corresponding to the global decrease in gene expression (for p-values demonstrating statistical significance of TAD and algebraic connectivity changes, see Table 3). A panoramic view of changes in TAD size and number is given for minimum TAD algebraic connectivities between 0.65 and 1, outside which the TAD structure devolves (Figure 4d). Figures 4b-c display the effect of trisomy on TADs with a minimum connectivity of 0.75, though analogous effects exist throughout the range of biological significance. 
 	
The TAD-scale analysis suggests that the genome compactifies, but does not inform us about the specifics of this change, for which we turn to network centrality analytics. Figure 5a shows a significant, uniform increase in degree centrality, decrease in betweenness centrality, increase in closeness centrality, and an insignificant erratic shift in eigenvector centrality in Chromosome 7 (for p-values demonstrating statistical significance of non-eigenvector centrality methods in all chromosomes, see Table 4). Briefly, an increase in degree centrality of a node in a Hi-C matrix implies that a DNA segment is connected to few other DNA segments, a decrease in betweenness centrality implies that a DNA segment is now less of a bottleneck, and an increase in closeness centrality implies that a DNA segment is - on average - spatially closer to other DNA segments (see table in Appendix B for an in-depth explanation). These centrality changes are uniform throughout the chromosome, and exist for all chromosomes with a low standard deviation (see Figure S2). The mean increase in degree centrality averages 19.64\% and the mean decrease in betweenness centrality averages 15.06\% (Figure 5b). Together, these centralities imply a uniform increase in the local of the genomic networks, corroborating the TAD-level analysis. The mean increase in closeness centrality averages 5.62\% and the eigenvector centrality change is both insignificant in magnitude (averaging $4.49*10^{(-14)}$\%) and inconsistent in direction, appearing like statistical noise (Figure 5c). The change in closeness centrality implies that the genomic networks are compactifying locally, but that the global distances remain intact, aligning with the eigenvector centrality, whose lack of change depicts a network whose nodes retain their relative importance. 

 \begin{figure}[htbp]
        \centering
               
         \begin{subfigure}[b]{0.7\textwidth}   
            \centering 
            \includegraphics[width=\textwidth]{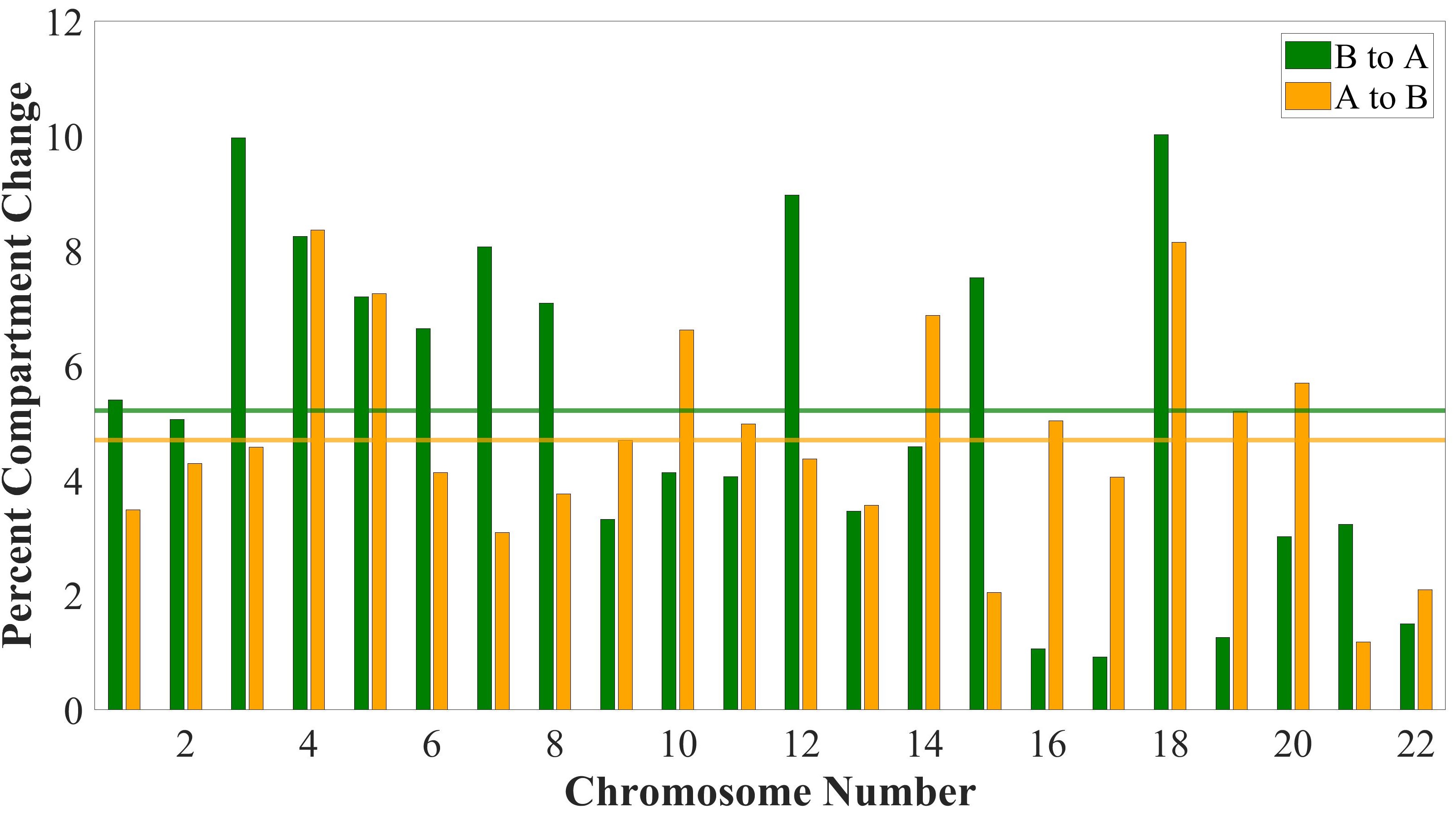}
            \caption[]%
            {{\small Percent change in A/B compartmentalization in each chromosome (an average of $4.70\%$ A-to-B, $5.21\%$ B-to-A)}} 
            \label{fig:3a}
        \end{subfigure}
         \quad
        \begin{subfigure}[b]{0.7\textwidth}
            \centering
            \includegraphics[width=\textwidth]{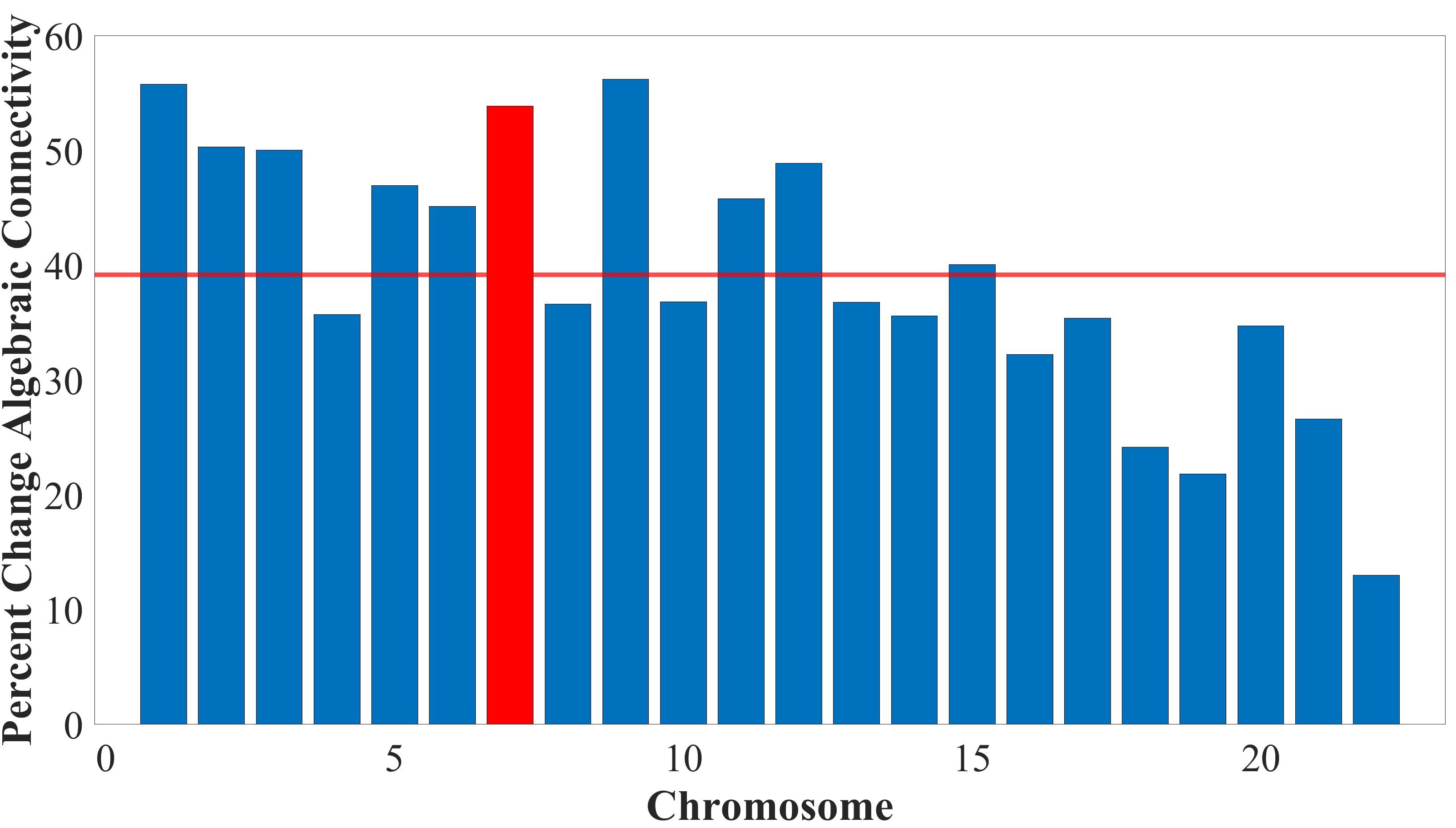}
            \caption[Network2]%
            {{\small Percent change in algebraic connectivity in each chromosome (an average increase of $39.17\%$)}}    
            \label{fig:3b}
        \end{subfigure}
        \caption{{\textbf{Major structural mechanisms influencing the observed changes in gene expression.} (a) Percentage of the chromosome changing from either A-to-B compartments (silencing) or B-to-A compartments (enabling) are similar (0.51 difference), though slightly more of the genome changes from B-to-A compartments than from A-to-B. (b) The percent change in the algebraic connectivity for each chromosome implies that the majority of each chromosome has become more physically connected.}} 
        \label{fig:3}
    \end{figure}

  \begin{figure}[htbp]
        \centering
        \begin{subfigure}[b]{0.6\textwidth}
            \centering
            \includegraphics[width=\textwidth]{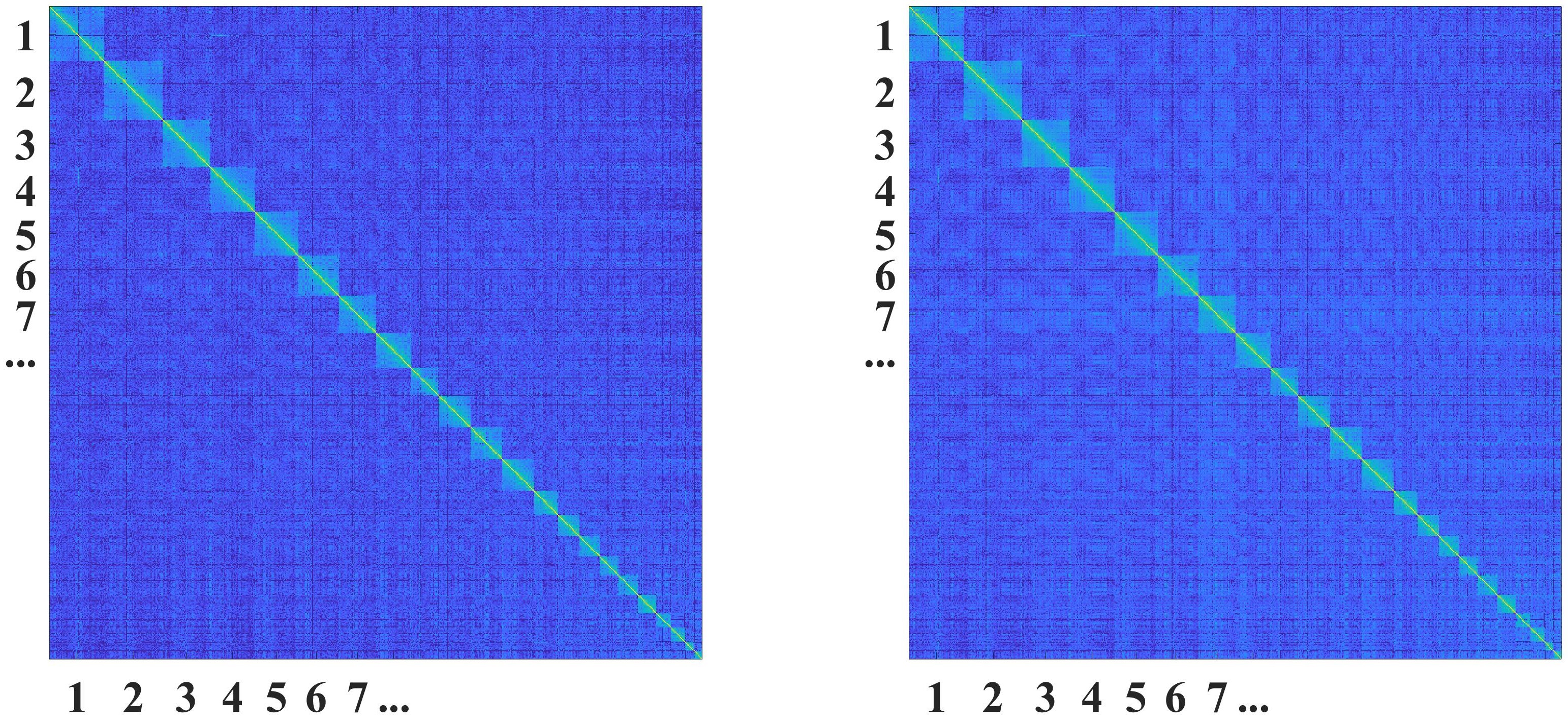}
            \caption[Network2]%
            {{\small Genome-wide Hi-C matrices}}    
            \label{fig:4a}
        \end{subfigure}
        \vskip\baselineskip
        \begin{subfigure}[b]{0.8\textwidth}  
            \centering 
            \includegraphics[width=\textwidth]{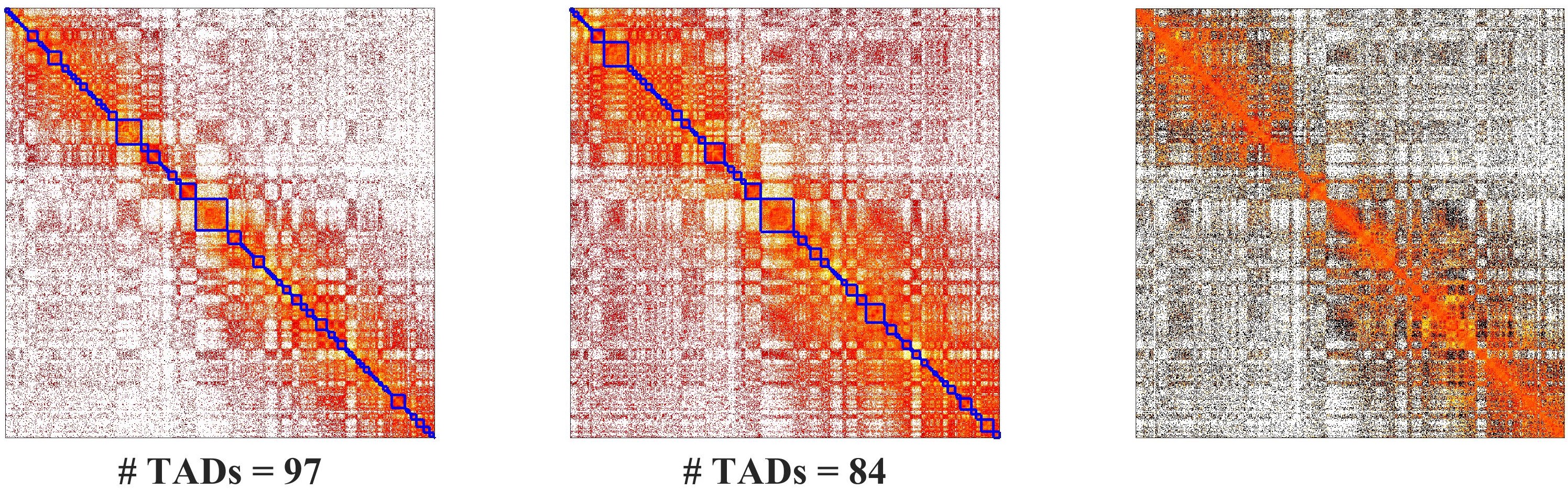}
            \caption[]%
            {{\small Sample chromosomal Hi-C matrices}}    
            \label{fig:4b}
        \end{subfigure}
        \vskip\baselineskip
        \begin{subfigure}[b]{0.475\textwidth}   
            \centering 
            \includegraphics[width=\textwidth]{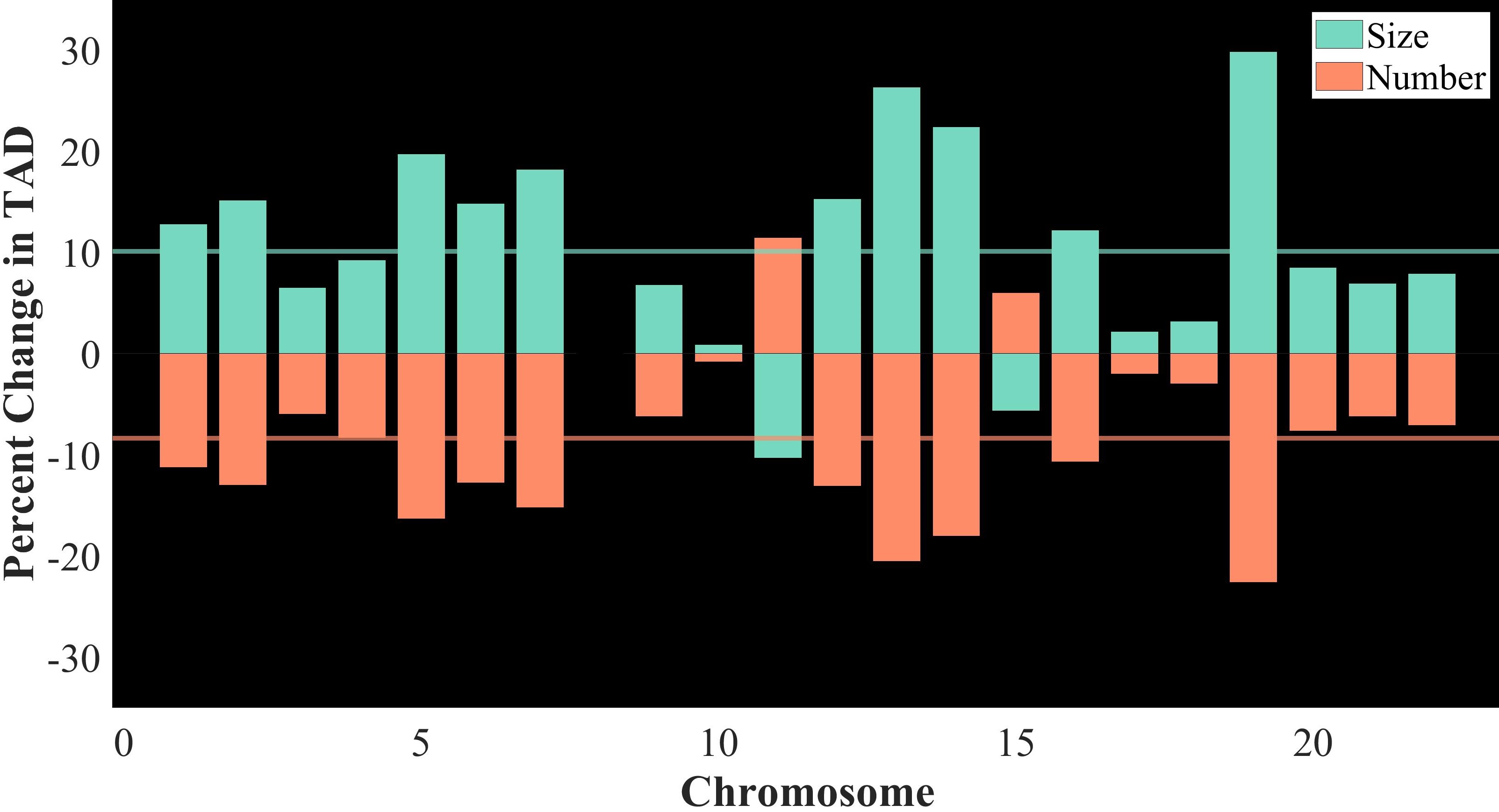}
            \caption[]%
            {{\small Percent change in TAD number (salmon, average $-9.59\%$ change) and size (turquoise, $11.35\%$ change)}}  
            \label{fig:4c}
        \end{subfigure}
        \quad
        \begin{subfigure}[b]{0.475\textwidth}   
            \centering 
            \includegraphics[width=\textwidth]{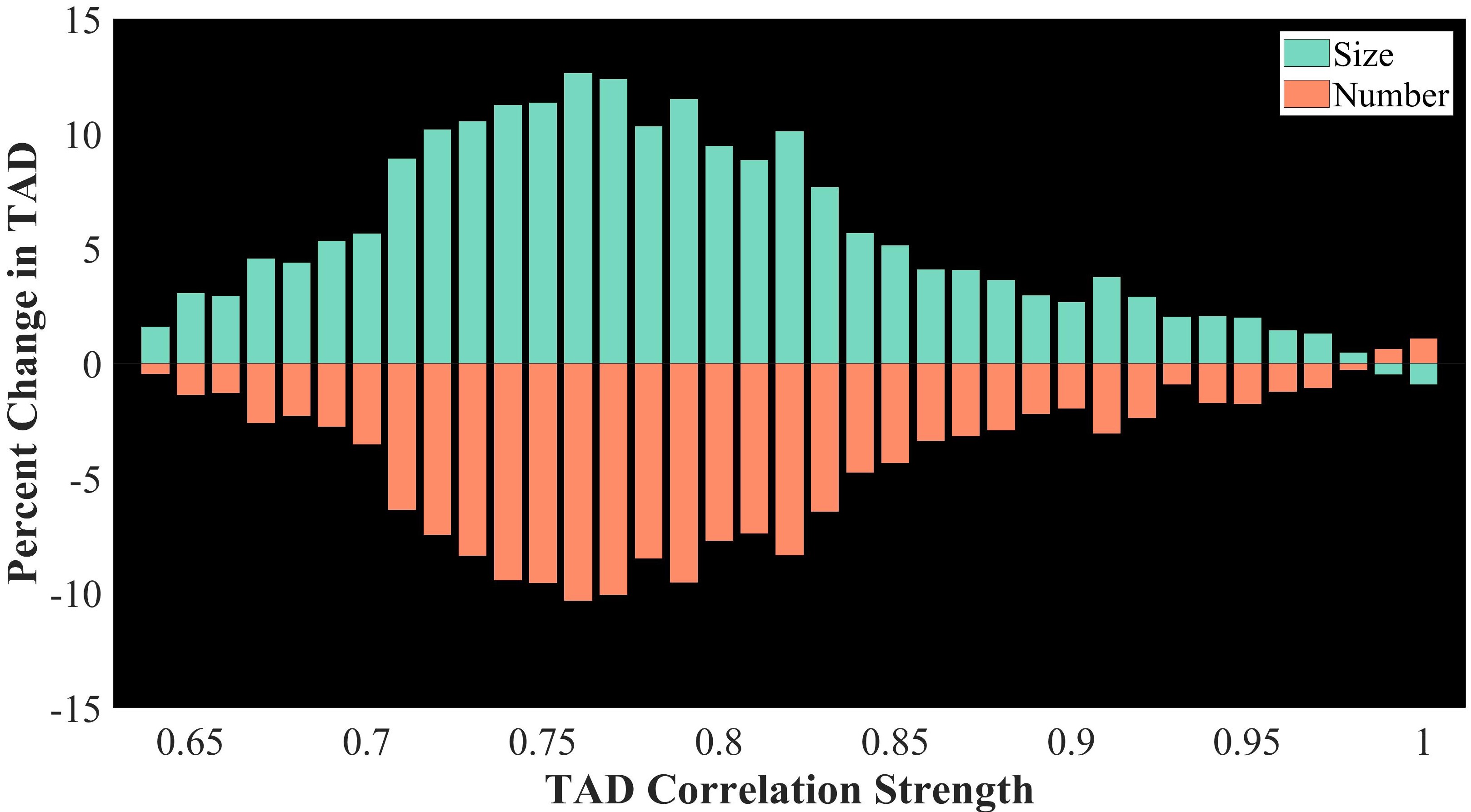}
            \caption[]%
            {{\small Percent change in TAD number and size for different thresholds)}} 
            \label{fig:4d}
        \end{subfigure}
        \label{fig:4}
        \caption{\textbf{Changes in TAD (high-connectivity area) size and number throughout the genome from healthy to trisomy-7 variant demonstrate an increase in global connectivity.} (a) Log-scaled Hi-C matrices for the entire genome, where brighter colors represent stronger connectivities. TAD squares match individual chromosomes, and on the right a strong cross pattern can be seen about Chromosome 7, indicating the trisomy. (b) The analog of figure (a) zoomed in to a single chromosome, where pixels represent 100kb in the genome. (c) The percent change in mean TAD size and number reveals that Chromosome 7 is well within the range of typical change, and that almost all chromosomes increase their mean TAD size and decrease in number of TADs. TADs are taken to require an interconnectivity of more than $0.77\%$ with a minimum size of 300kb. d) Averages of part (c) shown for different connectivity thresholds reveals that changes in TAD number and size are largest for the 'regions of biological significance' between 0.7 and 0.83.}
    \end{figure}

 \begin{figure}[htbp]
        \centering
        \begin{subfigure}[b]{0.7\textwidth}
            \centering
            \includegraphics[width=\textwidth]{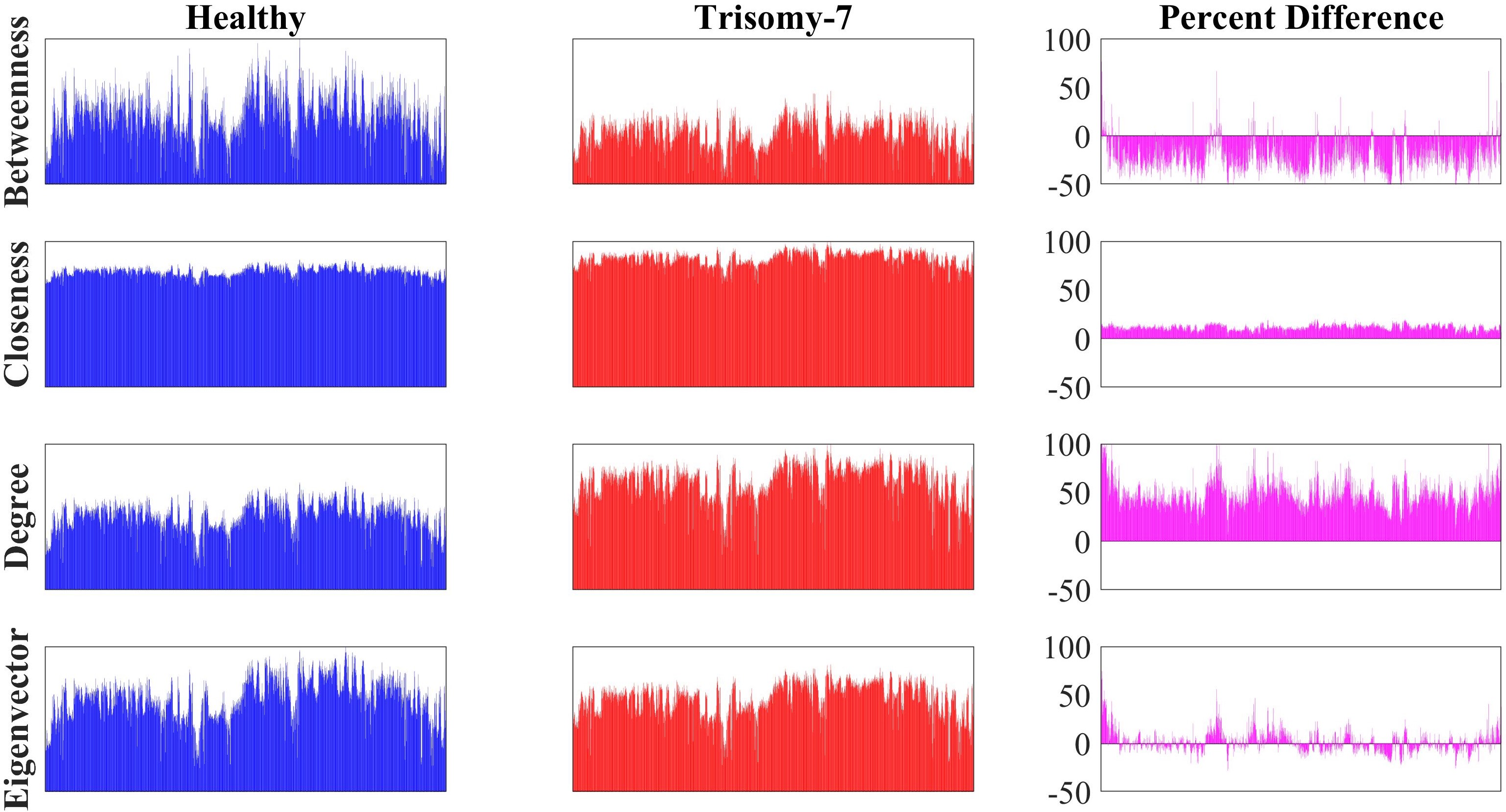}
            \caption[Network2]%
            {{\small Sample chromosome centralities}}    
            \label{fig:5a}
        \end{subfigure}
        \vskip\baselineskip

        \begin{subfigure}[b]{0.475\textwidth}   
            \centering 
            \includegraphics[width=\textwidth]{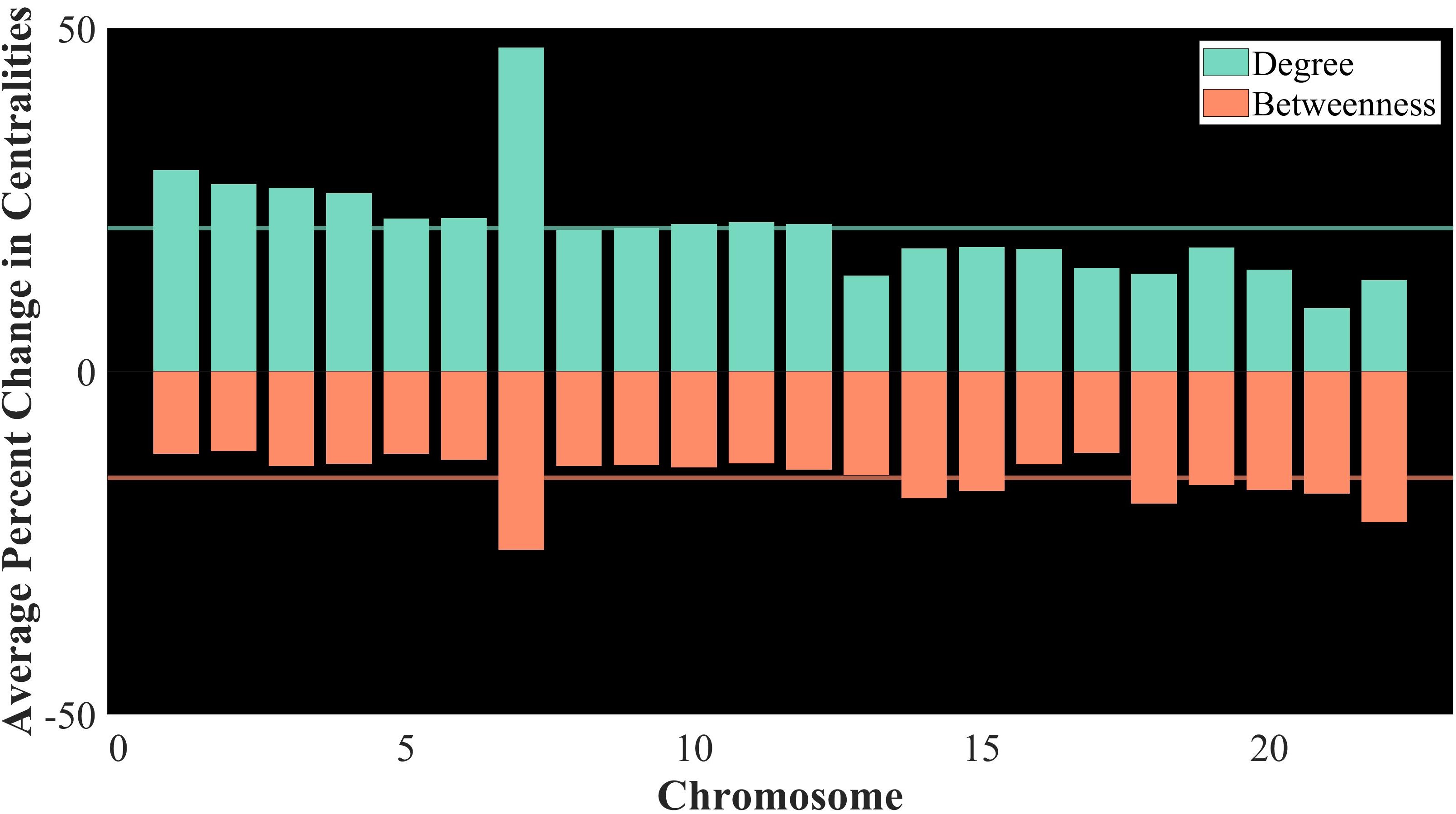}
            \caption[]%
            {{\small Percent change in betweenness (salmon, average: $15.62\%$ decrease) and degree (turquoise, average $20.89\%$ increase)   centralities}} 
            \label{fig:5b}
        \end{subfigure}
        \quad
        \begin{subfigure}[b]{0.475\textwidth}   
            \centering 
            \includegraphics[width=\textwidth]{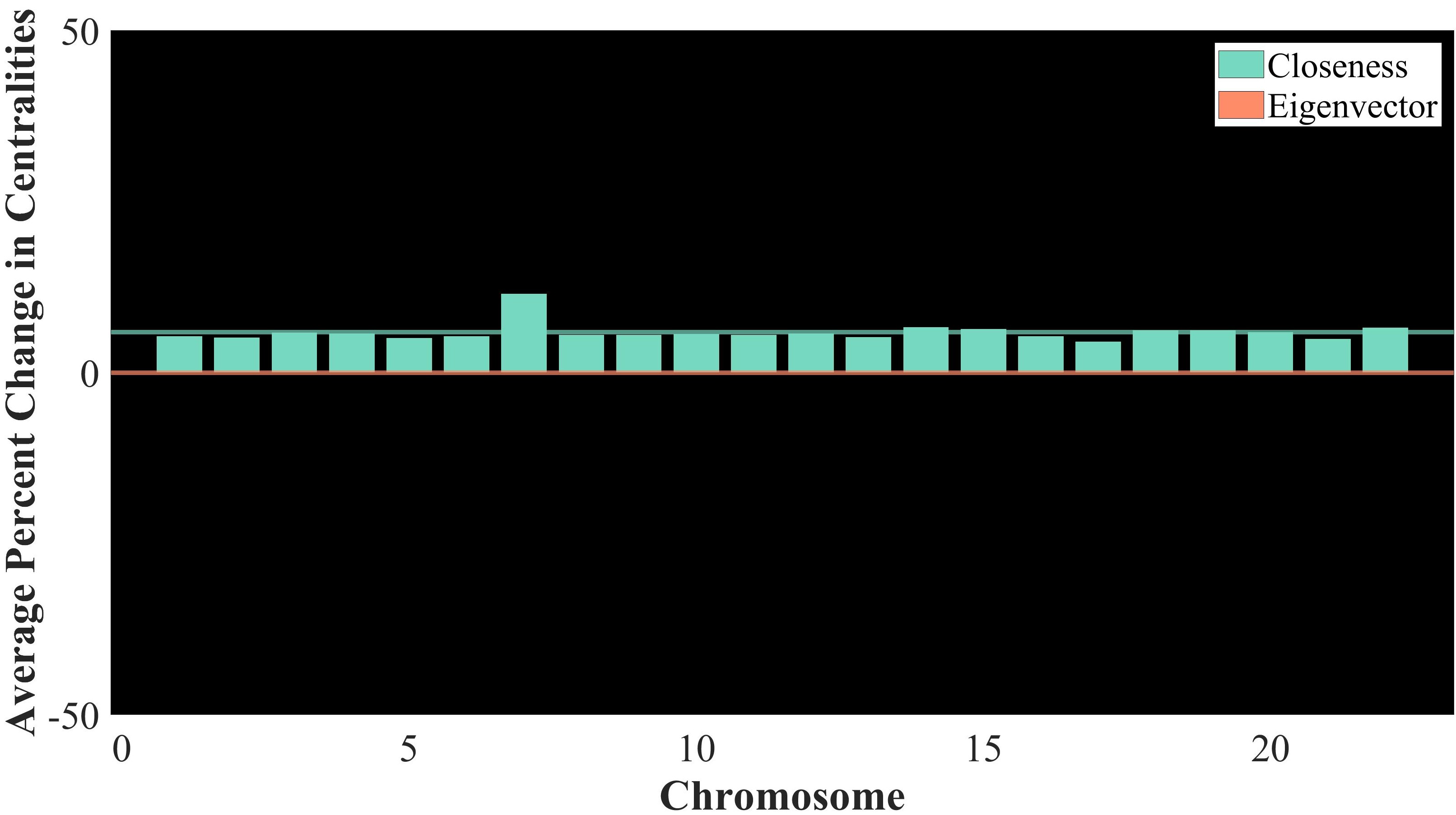}
            \caption[]%
            {{\small Percent change in eigenvector (salmon, average: $4.19 \times 10^{-14}\%$ decrease) and closeness (turquoise, average $5.89\%$ increase) centralities}}  
            \label{fig:5c}
        \end{subfigure}
        \caption{\textbf{Changes in centralities reveal an increase in local structural connections while maintaining the global architecture.} (a) A sample graph of centralities taken from Chromosome 7 depicts a stark decrease in betweenness centrality and increase in degree centrality, accompanied by a minor increase in closeness centrality and no clear pattern in the eigenvector centrality throughout the chromosome. (b) The mean percent change for degree (in turquoise) and betweenness (in salmon) centralities oppose in direction in all chromosomes. The global averages are summarized as horizontal lines in the corresponding colors. Chromosome 7 is the most significantly impacted by centrality alterations. (c) The mean percent change for closeness (in turquoise) and eigenvector (in salmon) centralities are shown for each chromosome. Chromosome 7 is the most significantly impacted by centrality alterations. The change in eigenvector centrality is insignificant - the mean line appears at zero. } 
        \label{fig:5}
    \end{figure}

\section{Analysis: Identifying and Characterizing the Jörmungandr Effect}

From Figure 2, we derived the existence of two different effects: Component 1, which causes large changes in gene expression in 8-10\% of each chromosome, and Component 2, which appears in at least 50\% of each chromosome but affects only minor changes in gene expression. From Figure 3, we identified Component 1 as the functional correlate of A/B compartment shifts. Where B-compartment shifts into A-compartment, there is significantly decreased connectivity. This yields a local spike upwards in betweenness centrality – as each segment is more likely to be a bottleneck, downward in degree centrality – as each segment is connected to less others, and downward in closeness centrality – as the network becomes less dense. Opposite structural shifts occur where A-compartment shifts into B-compartment. Thus, as the number of compartments shifting into A-compartment is similar to (if slightly higher than) those shifting into B-compartment, if Component 1 were the only effect present, we would expect occasional spikes up and down in centralities as described, with a slight bias toward the B-to-A changes. We would therefore expect the mean of each centrality measure to remain mostly unchanged, with perhaps a slight increase in betweenness and closeness centralities, and a corresponding decrease in degree centrality. We would also expect the global TAD structure to shift little, with the mean slightly biased toward an increase in number and decrease in size. However, we discover that all structural changes proceed strongly in the opposite direction in a uniform and global manner, meaning that the structural correlate of Component 2 exhibits strong changes in architectural metrics in the opposite direction of Component 1, which sum to overpower those of Component 1. 

To validate the biological relevance of Hi-C centrality metrics, we examine the consistency of these metrics when considered as physical measurements of the genomic architecture. The average closeness centrality increases by about 5.9\% from the HCEC to HCEC+7 cell lines, which yields a ratio of average closeness centrality of 1.059. Thus, we calculate the physical radius ratio (which is proportional to the inverse of closeness) to be 0.9443, or an average decrease in radius of about 5.57\%. Since the volume is proportional to the cube of the radius, the volume ratio is calculated to 0.842. The density is proportional to the inverse of the volume, and thus we calculate the density ratio of the aberrant to wild-type cells to be 1.188. Finally, since the trisomic cells have 47 chromosomes compared to 46 in the healthy cells, we multiply the density ratio by 47/46 to obtain 1.213, or an increase of 21.3\%, which is within the margin of machine error of our observed average degree centrality increase of 20.89\%. The consistency between measurements and their biological analogs validates the physical and biological significance of our selected metrics.

Following the vindication of centrality’s biological implications, we can assert the following about Component 2: (i) This effect is caused by the addition of a third copy of Chromosome 7. (ii) This effect is distinct from A/B compartmentalization shifts. (iii) This effect leads to a uniform global increase in connectivity, as is seen through the uniform changes in the forms of an increase in the mean size of TADs, reduction in the number of TADs, increase in degree centrality, and decrease in betweenness centrality. (iv) Component 2 does not change the overarching ‘shape’ of the physical structure of the genome, which is observed through the minor uniform increase in closeness centrality and lack of change in eigenvector centrality. (iv) Finally, the effect leads to a global, uniform decrease in the gene expression throughout the entire genome. 

The results support the existence of a novel mechanism for the global architectural shifts which yield the observed changes in gene expression - the Jörmungandr Effect. Named for the World Serpent, a colossal snake which surrounded the world by grasping its own tail in Norse mythology, the effect we propose is one which does exactly as its name suggests. The 3-D structure of the genome is tightened, resulting in a uniform, constant (5-6\%) decrease in radius and (20-21\%) increase in density, as well as significant increases in other metrics of connectivity that cause a similarly significant decrease in gene expression. To uniformly decrease the volume and radius, the chromosomes are physically condensed, which results in a decrease in gene expression due to a decrease in accessibility of the DNA. 

Recent results on trisomic cells in plants allude to a similar phenomenon. \cite{26,27} While they did not refer to structural changes, the gene expression changes they depict are analogous to the effects of our proposed mechanism. Their non-aneuploid chromosomes present a significant decrease in gene expression, and appear highly similar to our distribution plots. Their aneuploid chromosome plot presents an approximately 20\% median increase in gene expression, which is similar to our raw levels. Once we divide these levels by the aneuploid factor (3/2 in the case of trisomy), the results appear in the lower part of the range of other chromosomes, nearly identical to our distribution plot for the trisomic chromosome. Therefore, it may be the case that the Jörmungandr Effect is a general feature of aneuploidies.

\section{Conclusion}

We introduce a meta-analytic computational technique which aims to identify yet-uncharacterized biological effects. It begins by constructing a panoramic view of RNASeq for a range of thresholds as to preclude the loss of pertinent information from the selection of specific thresholds of analysis. We then account for known effects by renormalizing the data to counteract these effects. Thus, if all effects are accounted for, the data will be sufficiently close to the control state. Otherwise, we possess data filtered to differ from the control state via modification solely by previously unexplained effects. We name this method the Meta-analytic Operation of Threshold-independent Filtering (MOTiF) method. MOTiF can be applied in any case the net result of applying all effects is independent of their order (such as when all are additive, all are multiplicative, etc.). The state-of-the-art belief in gene expression is that all effects are multiplicative (or conversely divisive), and thus are ideal for the application of MOTiF.

As a proof of concept, we applied MOTiF to HCEC and HCEC+7 cell lines. First, we accounted for an extra copy of Chromosome 7, and then for the known A-B Compartment shifts that lead to large changes in gene expression in small portions of the genome, which in our data has an approximate equilibrium for increase and decrease of gene expression. After computationally undoing these effects in our data, it was still significantly different from the expected gene expression levels, which we were able to cross-validate with structural Hi-C data for the cell line. In tandem, these suggested a global condensation of the physical genomic architecture which resulted in the observed decrease in gene expression levels throughout all chromosomes.

It may seem surprising that such a significant, uniform decrease in gene expression and increase in structural connectivity has not been previously identified. A compelling answer is the typical thresholding analysis, wherein changes in gene expression are only considered above a factor of two, \cite{19,30} which entirely misses the Jörmungandr Effect. The benefit of providing a global view of the biology independent from arbitrarily chosen parameters is exactly that - freedom from subjectivity, focusing on the data itself to glean the juxtaposition of interlocking mechanisms. For these reasons, we suggest MOTiF as a state-of-the-art method for biological discovery.

We would like to stress the importance of caution and transparency when it comes to the methods of analysis and processing used on the data, which utilization of MOTiF facilitates. In the example of our data, using a minimum threshold for gene expression change above 1.3 (such as the commonly selected 2x threshold) would result in seeing only the A/B Compartmentalization and loss of the Jörmungandr Effect. Conversely, using a maximum threshold for gene expression change between 1 and 1.3 would result in seeing only the Jörmungandr Effect and loss of the A/B Compartmentalization. In either case, neglecting to normalize for the extra copy of Chromosome 7 would yield excessively pronounced results in this chromosome, which is not backed by the biological truth present in the data. 

Even after normalizing the data for the extra copy of Chromosome 7, the Jörmungandr Effect acts most powerfully upon the trisomic chromosome. As a result, we hypothesize that its purpose is to correct the excess gene expression brought about by the additional chromosome and ameliorate effects of this abnormal condition. However, based on the new expression levels, there is clearly some sort of overcompensation. The trisomic chromosome is now underexpressed, the average gene expression is below 91\% of normal levels, and the genome has condensed far beyond its previous healthy levels. We predict that this departure from the base homeostatic levels is one which is closely correlated to the tendency of aneuploidic cells to develop into carcinomatous tumors, perhaps resulting from the pressure to mutate away the disadvantages of aneuploidy. \cite{36}

One biological mechanism which aligns with the Jörmungandr Effect is DNA supercoiling, a winding of the strand around the central structural axis which can be taken as a measure of strain on the physical DNA strand. While other sorts of coiling and folding change the three-dimensional structure of the network, supercoiling retains the genome’s physical network structure (see Appendix A for more on supercoiling). This correlates with our observations, wherein the relative structure of the network (eigenvector centrality) is unchanged, while the network condenses (higher degree and closeness centralities) and becomes more tightly (lower betweenness centrality). Alternate mechanisms could include chemical and magnetic alterations or a physical tightening of an extra-nuclear membrane.

Future research should examine the connections between aneuploidy and additional supercoiling, as well as other potential candidates for the underlying biological mechanics of the Jörmungandr Effect. This effect must further be investigated in other trisomies, aneuploidies, and chromosomal aberrations (such as euploidies and chromosome-arm mutations) throughout a variety of organism. Then, it would be of utmost importance to understand how the Jörmungandr Effect relates to the propensity of aneuploidic cells for cancerous growth, as such a connection would open new horizons for early identification of the cancerous process. 


\bibliography{refs1}
\newpage
\section{APPENDIX A: Chromosomal Coiling and Condensation}

Each non-sex human cell contains two entire sets of 23 chromosomes, which becomes a total of around 6 billion base pairs, each 0.34 nanometers long, for about 2 meters of DNA per 10-100 micrometer long cell. The DNA’s negatively charged phosphate backbone is condensed via positively charged proteins called histones. The dense DNA-protein complex formed by coiling a DNA strand around histones is called chromatin, which consists of nucleosomes, segments consisting of 146 base pairs of DNA forming tight bonds around the eight histones proteins. \cite{37, 38} This process shortens the length of fiber by a factor of seven, which still leaves about 29 centimeters, far beyond the width of a cell. The chromatin is coiled into a 30-nanometer fiber (named for the diameter), then folded into 300-nanometer long loops, further compressed into a 250-nanometer-wide fiber, and finally coiled once again into the chromosome's dense chromatin DNA structure. \cite{39}

The amount of coiling needs to change frequently in cells - for making proteins, for DNA replication and cellular replication. For DNA replication and transcription, histones must be loosened to permit access to the DNA, which is done either enzymatically (by adding acetyl, methyl, or phosphates) or through chromatin remodeling complexes. \cite{40,41} Small segments of the DNA are uncoiled to transcribe them into mRNA, and this happens at thousands of points along a DNA strand per second, so the specific level of DNA density is in constant flux, while the average remains relatively constant. The exception to this is when cells replicate i.e. during mitosis or meiosis, and the level of DNA condensation is greatly increased  where DNA must be compacted and segregated to daughter cells. 

The intermediate step of coiling the chromatin into 10-(and then 30-)nanometer fibers is called DNA supercoiling, which is a winding of the strand around the axis and can be taken as a measure of strain on the physical DNA strand. This differs from the other kinds of coiling and folding, as while they change the three dimensional structure of the network, supercoiling retains the network structure. Most DNA is supercoiled twisted around itself in a right-hand fashion. Thus, coiling a strand further in the right-hand manner will further compactify the DNA and prevent access for replication and expression, while adding left-hand coils will loosen the tension and enable access. This is done through enzymes known as topoisomerases and gyrases among others, which can loosen or tighten the level of supercoiling to enable DNA replication or transcription. During DNA unwinding for transcription, additional (positive) supercoiling is induced ahead of the current area and thus negative supercoiling is correspondingly induced behind. \cite{42}

\section{APPENDIX B: Computational Methods}

The data used for this analysis first appeared in [31] and was processed as mentioned in [30]. As this paper is not about collecting new data, but rather about how to properly analyze pre-existing data, we focus only on a single model cell line with its associated functional and structural results.
 
In our Hi-C data, each chromosome has two corresponding matrices wherein each entry represents the number of times two 100-kilobase segments of DNA appear together in the ‘snapshots’ taken of the genome, one matrix for the healthy cells and the other for the trisomy-7 variant. The genome also has two such matrices, wherein instead of 100-kilobase segments, 1-megabase segments are used. The Hi-C matrix represents the number of counts where two segments of the genome are together. Biologically, if two parts of the genome are observed together more than a certain number of times, they must be close to each other, and can thus be considered ‘neighbors’. The ‘neighbor’ operation, both in the mathematical matrix model and in the physical biological model, possesses the following qualities: It is symmetric, such that a segment is connected to any segment which is connected to the first, it is reflexive, such that any segment is connected (particularly strongly) to itself, and it is not transitive, in that given three segments, if two of them are connected to the third, they are not necessarily connected to each other. Therefore, we can use our binarized Hi-C matrix as the adjacency matrix of a graph, wherein nodes represent specific segments of DNA and edges represent that these two nodes are physical neighbors.
 We normalized our Hi-C matrix using the Knight-Ruiz (KR) method, which turns a matrix into a doubly-stochastic orthonormal matrix, wherein each row and each column sum to the value of ‘1’. Many features that have been observed through Hi-C analysis, including A/B compartmentalization and topologically associating domains, are robust for the type of normalization, and is therefore a safe method of normalization which does not distort the data. \cite{32,43} To binarize the normalized Hi-C matrix, we used the standard value (which turns out to around 0.34 after Knight-Ruiz normalization), but for completeness a table containing average values of metrics used with different thresholds can be found in Appendix C (see Table 1), which demonstrates that any reasonable binarization value yields nearly identical results. 
Figure 3a was obtained by taking the first principal component of all chromosomes, with the exception of Chromosomes 4 and 5, for which the second principal component was taken. This is achieved by computing the covariance matrix and taking the eigenvectors which correspond to the largest (or in the case of Chromosomes 4 and 5, second largest) eigenvalue. In accordance with Lieberman, 2009, the DNA segments which correspond to positive values in these vectors are part of the highly expressed A-compartment, which is less densely connected. The areas which correspond to negative values are thus part of the highly condensed B-compartment, which possesses a lower rate of gene expression. Therefore, when comparing the same vectors between the HCEC and HCEC+7 cell lines, the areas which corresponded to positive values that changed to negative ones underwent A-to-B compartment shifting, while those that correspond to negative values that changed to positive ones shifting compartments from B-to-A. This subfigure depicts the percentage of each chromosome that shifted from A-to-B compartments, and those that went from B-to-A. In Figure 3b, we considered each chromosome’s Hi-C matrix to be a network and calculated the algebraic connectivity of each matrix, which is defined as the second smallest eigenvalue of the Normalized Laplacian of the original matrix. The Normalized Laplacian is defined as D-½(D-A)D-½ wherein A is the Hi-C matrix itself and D is a diagonal matrix wherein each entry is the sum of the corresponding row in A. The algebraic connectivities, which always range from 0 to 2, are a metric of how connected the chromosomes are, and thus how densely they are packed. 

In Figures 4c-d the TADs were computed using the 4DNvestigator kit, providing the program with a minimum number of blocks to be considered a TAD, a minimum connectivity which is necessary to be considered a TAD (for which 0.75 was used for 4c and a range of values between 0.65 and 1 was used for 4d) and an Observed/Expected (O/E) normalization of the Hi-C matrix. \cite{44}

In Figure 5, considering the Hi-C as an adjacency matrix, we computed four different methods of centralities for each segment of each chromosome, displaying the vector resulting from doing so for each ‘node’ of a single chromosome in Figure 5a for the wild-type and the trisomic variants along with percent differences. In 5b and 5c, we displayed only the mean values for percent differences in each centrality measure for each chromosome between the HCEC and HCEC+7 cell lines. Please refer to the table below for the formulas, definitions, and biological significance in Hi-C for each of the centralities. For reference, $e_{i,j}$ denotes an edge between vertex $i$ and vertex $j$, being ’1’ if such an edge exists and ’0’ if one does not. $d_{i,j}$ denotes the geodesic distance between nodes $i$ and $j$, i.e. the minimum path length from node $i$ to node $j$. $g_{j,k}$  denotes the number of geodesics (paths with the shortest distance) between nodes $j$ and $k$, and $g_{j,k}(i)$ denotes the number of such geodesics that pass through node $i$. Finally, $\lambda$ denotes the largest eigenvalue of the adjacency (Hi-C) matrix.

\begin{center}
\begin{tabular}{|M{2cm}|M{4cm}|M{3cm}|M{6cm}| }
 \hline \vspace{2mm} Centrality \vspace{1mm} & \vspace{2mm} Definition \vspace{1mm} & \vspace{2mm} Equation \vspace{1mm} & \vspace{2mm} Biological Significance \vspace{1mm} \\ 
\hline \hline Betweenness & \vspace{2mm} The betweenness centrality of a node $n$ is defined as the sum (over all pairs of nodes $u$ and $v$) of ratios between geodisics (shortest paths between $u$ and $v$) which pass through $n$ and all $u$-$v$ geodisics. \vspace{2mm} & $c^b(i)= \sum_{j<k}  \frac{g_{j,k}(i)}{g_{j,k}}$ & Biologically, an increase (\textbf{decrease}) in betweenness centrality of a node in the Hi-C matrix implies that a given stretch of DNA is now more (\textbf{less}) of a bottleneck, insofar as a removal of that DNA segment would distance more (\textbf{less}) segments from each other. \\ 

\hline \hline Closeness & \vspace{2mm} The closeness centrality of a node n is defined as the inverse of the sum of distances from $n$ to all other nodes. A higher closeness centrality implies a shorter diameter. \vspace{2mm} & $c^c(i)=\frac{1}{\sum_{j:j\neq i} d_{i,j}}$ & Biologically, an increase (\textbf{decrease}) in closeness centrality of a node in the Hi-C matrix implies that a given stretch of DNA is now - on average - spatially closer to (\textbf{further from}) other DNA segments.  \\ 

\hline \hline Degree & \vspace{2mm} The degree centrality of a node $n$ is defined as the number of nodes to which it is adjacent. A higher degree centrality implies a higher density of nodes. \vspace{2mm} & $c^d (i)= \sum_{j:j\neq i} e_{i,j}$ & Biologically, an increase (\textbf{decrease}) in degree centrality of a node in the Hi-C matrix implies that a given stretch of DNA is now connected to - more highly topologically related with - a greater (\textbf{lesser}) number of other DNA segments.   \\ 

\hline \hline Eigenvector & The eigenvector centrality of a node $n$ is defined as the weighted sum of the eigenvector centrality of its neighbors. & $c^e(i)=\frac{1}{\lambda} \sum_{j:j\neq i} e_{i,j} c^e(j)$ & \vspace{2mm} Biologically, an increase (\textbf{decrease}) in eigenvector centrality of a node in the Hi-C matrix implies that a given stretch of DNA becomes more (\textbf{less}) influential within the network, i.e. that it is now adjacent with DNA segments that are also more (\textbf{less}) influential, and implies a change in the broad, global shape of the architecture. \vspace{2mm}  \\ 

\hline

\end{tabular}
\end{center}
\noindent

\newpage 

\section{APPENDIX C: Additional Measurements and Statistical Significance}

\begin{center}
\[\begin{array}{|M{2cm}|M{2cm}|M{2cm}|M{2cm}|M{2.75cm}|M{2cm}|}
\hline \vspace{2mm} Hi-C Normalized Cutoff \vspace{1mm} & \vspace{2mm} Betweenness Centrality Change \vspace{1mm} & \vspace{2mm} Closeness Centrality
Change \vspace{1mm} & \vspace{2mm} Degree Centrality
Change \vspace{1mm} & \vspace{2mm} Eigenvector Centrality
Change \vspace{1mm}& \vspace{2mm} Algebraic Connectivity
Change \vspace{1mm} \\ 
\hline \vspace{1mm} 0.3 \vspace{1mm} &	-15.560\% &	5.8903\% &	20.8907\% &	$-4.188 \times 10^{-14}$\% &	39.174\% \\
\hline \vspace{1mm} 0.4 \vspace{1mm} &	-15.560\% &	5.8903\% &	20.8909\% &	$-3.280 \times 10^{-14}$\% &	39.175\% \\
\hline \vspace{1mm} 0.5 \vspace{1mm} &	-15.560\% &	5.8904\% &	20.8921\% &	$-4.895 \times 10^{-14}$\% &	39.175\% \\
\hline \vspace{1mm} 0.6 \vspace{1mm} &	-15.584\% &	5.9011\% &	20.9271\% &	$-8.881 \times 10^{-14}$\%	& 39.210\% \\
\hline \vspace{1mm} 0.7 \vspace{1mm} &	-16.183\% &	6.1836\% &	22.0427\% &	$-3.785 \times 10^{-14}$\% &	40.539\% \\
\hline \vspace{1mm} 0.8 \vspace{1mm} &	-17.835\% &	7.0874\% &	27.0761\% &	$-8.276 \times 10^{-14}$\%&	46.565\% \\
\hline \vspace{1mm} 0.9 \vspace{1mm} &	-15.254\% &	6.3082\% &	28.4282\% &	$-1.867 \times 10^{-14}$\% &	49.531\% \\
\hline \vspace{1mm} 1.0 \vspace{1mm} &	-9.803\% &	4.1234\% &	22.0569\% &	$4.188 \times 10^{-14}$\%&	43.608\% \\
\hline

\end{array}\]
\end{center}
 \noindent					
\textbf{Table 1 - Centralities and Algebraic Connectivity for Different Hi-C Neighborhood Matrix Cutoffs.} The cutoff point for creation of the neighborhood matrix is typically taken to be $0.34$ after KR normalization. For completeness, percent change in centralities and algebraic connectivity between the HCEC and HCEC+7 cell lines are provided for all cutoff points between $0.3$ and $1$, above and below which the neighborhood matrix devolves beyond the point of analysis. We see that as expected, within a reasonable cutoff zone of $0.3$ and $0.5$, numerical results are identical, and are nearly identical at $0.6$. The same effect is exhibited for all higher cutoff points, although the magnitudes of change undergo slight perturbation, as high cutoff points induce unstable neighborhood matrices.

\begin{center}
\[\begin{array}{|M{3cm}|M{5cm}|M{3cm}|}
\hline \vspace{2mm} Chromosome \vspace{1mm} & \vspace{2mm} P-value for Expression Change \vspace{1mm} & \vspace{2mm} Significant? \vspace{1mm}  \\ 

\hline \vspace{1mm} 1 \vspace{1mm} &	$1.17 \times 10^{-6}$ &	Yes \\
\hline \vspace{1mm} 2 \vspace{1mm} &	0.678919 &	No \\
\hline \vspace{1mm} 3 \vspace{1mm} &	$9.00 \times 10^{-10}$ &	Yes \\
\hline \vspace{1mm} 4 \vspace{1mm} &	$9.32 \times 10^{-5}$ &	Yes \\ 
\hline \vspace{1mm} 5 \vspace{1mm} &	$2.23 \times 10^{-5}$ &	Yes \\ 
\hline \vspace{1mm} 6 \vspace{1mm} & 	0.009556 &	Yes \\
\hline \vspace{1mm} 7 \vspace{1mm} &	$8.98 \times 10^{-5}$ &	Yes \\
\hline \vspace{1mm} 8 \vspace{1mm} &	$8.90 \times 10^{-7}$ &	Yes \\ 
\hline \vspace{1mm} 9 \vspace{1mm} &	$2.67 \times 10^{-8}$ &	Yes \\
\hline \vspace{1mm} 10 \vspace{1mm} &	0.412481 &	No \\
\hline \vspace{1mm} 11 \vspace{1mm} &	0.001456 &	Yes \\ 
\hline \vspace{1mm} 12 \vspace{1mm} &	0.006151 &	Yes \\
\hline \vspace{1mm} 13 \vspace{1mm} &	0.092667 &	No \\
\hline \vspace{1mm} 14 \vspace{1mm} &	$1.78 \times 10^{-7}$ &	Yes \\
\hline \vspace{1mm} 15 \vspace{1mm} &	0.022471 &	Yes \\
\hline \vspace{1mm} 16 \vspace{1mm} &	$8.50 \times 10^{-5}$ &	Yes \\
\hline \vspace{1mm} 17 \vspace{1mm} &	0.222389 &	No \\
\hline \vspace{1mm} 18 \vspace{1mm} &	0.000258 &	Yes \\
\hline \vspace{1mm} 19 \vspace{1mm} &	0.026263 &	Yes \\
\hline \vspace{1mm} 20 \vspace{1mm} &	0.000421 &	Yes \\
\hline \vspace{1mm} 21 \vspace{1mm} &	0.660455 &	No \\
\hline \vspace{1mm} 22 \vspace{1mm} &	0.62982 &	No \\

\hline 
\end{array}\]
\end{center}
  \noindent					
\textbf{Table 2 - Statistical Significance of Change in Gene Expression.}
When taken collectively, the change in gene expression levels from the HCEC to HCEC+7 cell lines are highly statistically significant. For completeness, we provide statistical significance of the changes in gene expression levels for each chromosome. Of the 23 chromosomes, 17 exhibit a statistically significant change in gene expression levels between the two cell lines.

\begin{center}
\[\begin{array}{|M{4.5cm}|M{3.5cm}|M{3.5cm}|}
\hline \vspace{2mm} Metric \vspace{1mm} & \vspace{2mm} P-value\vspace{1mm} & \vspace{2mm} Significant? \vspace{1mm}  \\ 

\hline  \vspace{1mm}  Mean TAD Size \vspace{1mm} & $3.92\times 10^{-6}$ & Yes \\
\hline  \vspace{1mm} TAD Number \vspace{1mm} &	$2.13 \times 10^{-5}$  &	  Yes \\
\hline  \vspace{1mm}  Algebraic Connectivity  \vspace{1mm} &	$9.88 \times 10^{-15}$ &	  Yes \\

\hline 
\end{array}\]
\end{center}
 \noindent	
\textbf{Table 3 - Statistical Significance of Connectivity Metrics.}
The non-centrality connectivity metrics all exhibit a highly statistically significant change from the HCEC to HCEC+7 cell lines.
\\ \ \\

\begin{center}
\[\begin{array}{|M{2cm}|M{2.5cm}|M{2.5cm}|M{2.5cm}|M{2.5cm}|}
\hline \vspace{1mm} \vspace{4mm} Chromosome \vspace{1mm} & \vspace{2mm} P-value for Betweenness Centrality \vspace{1mm} & \vspace{2mm} P-value for Degree Centrality  \vspace{1mm} & \vspace{2mm} P-value for Closeness Centrality \vspace{1mm} & \vspace{2mm} P-value for Eigenvector Centrality \vspace{1mm}  \\ 

\hline \vspace{1mm} 1 \vspace{1mm} &	$2.66 \times 10^{-174}$ &	0 &	0 &	1 \\
\hline \vspace{1mm} 2 \vspace{1mm} &	$8.58 \times 10^{-231}$ &	0 &	0 &	1 \\
\hline \vspace{1mm} 3 \vspace{1mm} &	$5.78 \times 10^{-198}$ &	0 &	0 &	1 \\
\hline \vspace{1mm} 4 \vspace{1mm} &	$8.78 \times 10^{-120}$ &	0 &	0 &	1 \\
\hline \vspace{1mm} 5 \vspace{1mm} &	$6.41 \times 10^{-139}$ &	0 &	0 &	1 \\
\hline \vspace{1mm} 6 \vspace{1mm} &	$1.20 \times 10^{-164}$ &	0 &	0 &	1 \\
\hline \vspace{1mm} 7 \vspace{1mm} &	0 &	0 &	0 &	1 \\
\hline \vspace{1mm} 8 \vspace{1mm} &	$4.95 \times 10^{-140}$ &	0 &	0 &	1 \\
\hline \vspace{1mm} 9 \vspace{1mm} &	$3.09 \times 10^{-108}$ &	0 &	0 &	1 \\
\hline \vspace{1mm} 10 \vspace{1mm} &	$5.04 \times 10^{-146}$ &	0 &	0 &	1 \\
\hline \vspace{1mm} 11 \vspace{1mm} &	$1.09 \times 10^{-134}$ &	0 &	0 &	1 \\
\hline \vspace{1mm} 12 \vspace{1mm} &	$4.87 \times 10^{-147}$ &	0 &	0 &	1 \\
\hline \vspace{1mm} 13 \vspace{1mm} &	$7.99 \times 10^{-130}$ &	0 &	0 &	1 \\
\hline \vspace{1mm} 14 \vspace{1mm} &	$1.51 \times 10^{-124}$ &	$6.59 \times 10^{-306}$ &	0 &	1 \\
\hline \vspace{1mm} 15 \vspace{1mm} &	$4.62 \times 10^{-102}$ &	$2.31 \times 10^{-291}$ &	$3.26 \times 10^{-301}$ &	1 \\
\hline \vspace{1mm} 16 \vspace{1mm} &	$8.53 \times 10^{-72}$ &	$0.87 \times 10^{-273}$ &	$5.00 \times 10^{-298}$ &	1 \\
\hline \vspace{1mm} 17 \vspace{1mm} &	$4.98 \times 10^{-65}$ &	$1.90 \times 10^{-235}$ &	$3.25 \times 10^{-239}$ &	1 \\
\hline \vspace{1mm} 18 \vspace{1mm} &	$3.70 \times 10^{-120}$ &	$3.44 \times 10^{-246}$ &	$6.76 \times 10^{-259}$ &	1 \\
\hline \vspace{1mm} 19 \vspace{1mm} &	$3.39 \times 10^{-78}$ &	$4.30 \times 10^{-194}$ &	$6.27 \times 10^{-203}$ &	1 \\
\hline \vspace{1mm} 20 \vspace{1mm} &	$4.36 \times 10^{-103}$ &	$7.18 \times 10^{-221}$ &	$7.84 \times 10^{-228}$ &	1 \\
\hline \vspace{1mm} 21 \vspace{1mm} &	$1.27 \times 10^{-39}$ &	$1.28 \times 10^{-95}$ &	$6.17 \times 10^{-92}$ &	1 \\
\hline \vspace{1mm} 22 \vspace{1mm} &	$1.75 \times 10^{-70}$ &	$1.68 \times 10^{-109}$ &	$1.20 \times 10^{-112}$ &	1 \\

\hline

\end{array}\]
\end{center}
 \noindent	
\textbf{Table 4 - Statistical Significance of Centralities.}
All centrality metrics other than eigenvector centrality exhibit a highly statistically significant change from HCEC to HCEC+7 cell lines in all chromosomes, and eigenvector centrality is entirely indistinguishable in all chromosomes.

\end{document}